\documentclass{amsart}

\usepackage{amsmath,amssymb,amsthm}
\usepackage[nobysame]{amsrefs}
\usepackage[ascii]{inputenc}
\usepackage{graphicx}
\usepackage[caption=false]{subfig}
\usepackage{bbm}
\usepackage{indentfirst}
\usepackage{mathrsfs}
\usepackage{fourier}
\usepackage[pdftex,bookmarks=false,colorlinks=true,linkcolor=blue,citecolor=blue,filecolor=black,urlcolor=blue]{hyperref}
\usepackage[margin=1.2in]{geometry}

\theoremstyle{plain}

\DeclareMathOperator{\tr}{tr}
\DeclareMathOperator{\minmod}{minmod}

\providecommand{\ud}{\,\mathrm{d}}
\providecommand{\abs}[1]{\lvert#1\rvert}
\providecommand{\average}[1]{\left\langle#1\right\rangle}
\providecommand{\bra}[1]{\langle#1\rvert}
\providecommand{\ket}[1]{\lvert#1\rangle}
\providecommand{\eff}{\mathrm{eff}}
\providecommand{\FD}{\mathrm{FD}}
\providecommand{\bd}[1]{\boldsymbol{#1}}
\providecommand{\RR}{\mathbb{R}}

\providecommand{\ZZ}{\mathbb{Z}}
\providecommand{\CC}{\mathbb{C}}
\providecommand{\Or}{\mathcal{O}}

\IfFileExists{mathabx.sty}%
  {\DeclareFontFamily{U}{mathx}{\hyphenchar\font45}%
   \DeclareFontShape{U}{mathx}{m}{n}{<->mathx10}{}%
   \DeclareSymbolFont{mathx}{U}{mathx}{m}{n}%
   \DeclareFontSubstitution{U}{mathx}{m}{n}%
   \DeclareMathAccent{\widebar}{0}{mathx}{"73}%
}{%
  \PackageWarning{mathabx}{%
    Package mathabx not available, therefore\MessageBreak substituting
    widebar with overline\MessageBreak }%
  \newcommand{\widebar}[1]{\overline{#1}}%
}
\providecommand{\wt}[1]{\tilde{#1}}
\providecommand{\wh}[1]{\widehat{#1}}
\providecommand{\wb}[1]{\widebar{#1}}

\title[Numerical scheme for a spatially inhomogeneous matrix-valued
Boltzmann equation]{Numerical scheme for a spatially inhomogeneous \\
  matrix-valued quantum Boltzmann equation}

\author{Jianfeng Lu}
\address{Departments of Mathematics, Physics, and Chemistry, Duke University, Box 90320, Durham, NC 27708 USA}
\email{jianfeng@math.duke.edu}

\author{Christian B. Mendl} 
\address{Mathematics Department, Technische Universit\"at M\"unchen, Boltzmannstra{\ss}e 3, 85747 Garching bei M\"unchen, Germany}
\email{mendl@ma.tum.de}

\date{\today} 

\thanks{We would like to thank Jingwei Hu, Lorenzo Pareschi and Herbert Spohn for
  helpful discussions. The work of J.L. was supported in part by the
  Alfred P.~Sloan Foundation and the National Science Foundation under
  award DMS-1312659. C.M. would like to thank the warm hospitality of
  the Mathematics Department at Duke University where part of the work
  was done, as well as support from DFG}
\begin{document}

\begin{abstract}
We develop an efficient algorithm for a spatially inhomogeneous matrix-valued quantum Boltzmann equation derived from the Hubbard model. The distribution functions are $2 \times 2$ matrix-valued to accommodate the spin degree of freedom, and the scalar quantum Boltzmann equation is recovered as special case when all matrices are proportional to the identity. We use Fourier discretization and fast Fourier transform to efficiently evaluate 
the collision kernel with spectral accuracy, and numerically investigate periodic, Dirichlet and Maxwell boundary conditions. Model simulations quantify the convergence to local and global thermal equilibrium.
\end{abstract}

\maketitle

\section{Introduction}

Boltzmann's kinetic theory is widely used to describe the dynamics of rarified gases. Immediately after the discovery of quantum mechanics, a modification of the classical Boltzmann equation to take quantum interactions into account has been proposed by Nordheim \cite{Nordheim1928} (with a more systematic derivation by Peierls \cite{Peierls:29}), and by Uehling and Uhlenbeck \cites{UehlingUhlenbeck1933, Uehling1934}. With the inclusion of quantum mechanical effects into the collision operator, the quantum Boltzmann equation has many applications, e.g., for the kinetic description of Bose-Einstein condensation \cites{SemikozTkachev:95, Spohn:10}, spintronics and decoherence theory in quantum computing
\cites{Spintronics2001, ElHajj2014, PossannerNegulescu2011, VacchiniHornberger2009}, or kinetic modeling of semiconductor devices \cite{MarkowichRinghoferSchmeiser:90}. 

In recent works, starting from the Hubbard model in the weak interaction limit, a matrix-valued Boltzmann equation has been derived \cites{BoltzmannHubbard2012, BoltzmannNonintegrable2013, DerivationBoltzmann2013} for the spatially homogeneous setting without advection term. To describe spatially inhomogeneous systems, one combines the Boltzmann transport equation with the collision term derived in \cite{DerivationBoltzmann2013}:
\begin{equation}\label{eq:boltzmann}
  \partial_t W + v_x\,\partial_{x} W = \mathcal{C}[W] - i [ \vec{B} \cdot \vec{\sigma}, W],
\end{equation}
where the state variable $W$ is the Wigner distribution of the
spin-density matrix, $\vec{B}$ is an external magnetic field,
$\vec{\sigma}$ the Pauli matrices, and the collision term
$\mathcal{C}[W]$ will be specified below in
Section~\ref{sec:boltzmann}. We emphasize that while the form of the
equation follows the usual quantum Boltzmann equation, the collision
term is quite different (which is systematically derived from a
many-body quantum mechanics model), and the matrix-valued $W$
distinguishes the equation from the usual kinetic equations.

The focus of this paper is devising an efficient algorithm for solving \eqref{eq:boltzmann}. Our goal in this work is twofold: First, we would like to develop a numerical scheme that systematically converges to the true solution; for that purpose, we use a spectral method in the velocity variable. The collision operator, albeit much more complicated than for the usual classical or scalar quantum Boltzmann equation, can be efficiently calculated using Carleman representation and fast Fourier transforms. Second, we want to investigate non-trivial boundary conditions, like Dirichlet and Maxwell boundary conditions, and the effect of external magnetic fields. These developments should lead to a better understanding of the physics modeled by these
equations.

\smallskip

Kinetic equations are traditionally solved by Monte Carlo methods
(also known as particle methods). In recent years, the development of
efficient real space or Fourier space methods to solve Boltzmann
equations has been a very active research area. In particular, the
line of research initiated by \cite{MouhotPareschi:06} and further
developed in \cites{FilbertHuJin:12, HuYing:12} is especially relevant
for our approach. The paper \cite{MouhotPareschi:06} proposed a fast
algorithm for computing the Boltzmann collision kernel based on
Fourier discretization, and \cite{HuYing:12} further improved the
efficiency of the algorithm. The method we develop in this work for
the collision operator of the matrix-valued Boltzmann equation is
closely related, albeit with some differences: (a) Since we are
dealing with collision terms originating from quantum mechanics, the
microscopic energy is not necessarily conserved (see the effective
Hamiltonian in Eq.~\eqref{eq:Heff} below); some new ideas are required
to treat the resulting terms. (b) To evaluate double convolutions
exactly using fast Fourier transforms, we use a double padding
approach to avoid aliasing issues. As a result, while maintaining the
spectral accuracy, the scheme also nicely respects the conservation
law of the continuous equation. Further details can be found in
Section~\ref{sec:collision}.

A numerical algorithm for the spatially homogeneous matrix-valued
Boltzmann equation in one dimension was considered before in
\cites{BoltzmannHubbard2012, BoltzmannNonintegrable2013}, which
calculates the collision term directly using numerical quadrature. A
subsequent work by one of the authors \cite{Mendl2013} considers a
lattice Boltzmann method (LBM) for the spatially inhomogeneous equation
with periodic boundary conditions,
which can be understood as a discrete velocity method with very few
velocity grid points. Due to the small number of grid points, the accuracy of
the numerical result compared to the original equation is not
guaranteed. In contrast to that, the method proposed here
systematically approximates the original equation as we refine the grid. 

Considering the \emph{matrix-valued} Boltzmann equation, it is useful to represent the
spin-density Wigner distribution in the basis of Pauli matrices. In
fact, the formulas for the collision terms are more compact in the new
representation, which might be of independent interest for understanding the physics and mathematics of the equation. 

The rest of the paper is organized as follows. We introduce the
spatially inhomogeneous matrix-valued Boltzmann equation and its
associated boundary conditions in Section~\ref{sec:boltzmann}. We will
focus on the development of the fast algorithm for the collision
operator in Section~\ref{sec:collision}. To deal with the spatial
degree of freedom, we use the finite volume method and a parallel
implementation based on MPI; this is discussed in
Section~\ref{sec:parallel}. We show some numerical results for
validating the algorithm and for exploring interesting physical
phenomena in Section~\ref{sec:numerics}. Finally we wrap up the paper
with some conclusive remarks in Section~\ref{sec:conclusion}.

\section{The spatially inhomogeneous matrix-valued Boltzmann equation}
\label{sec:boltzmann}

The starting point for the derivation \cite{DerivationBoltzmann2013}
of the matrix-valued Boltzmann equation is the Hubbard model with a
weak pair potential $\lambda V$ such that $0 < \lambda \ll
1$. Consider a spin-$\frac{1}{2}$ Fermi field with annihilation operators
$a_{s}(x)$, $x \in \ZZ^d$, $s \in \{\uparrow, \downarrow
\}$, obeying the anti-commutation relations
\begin{equation*}
  \big\{ a_{s}(x)^{\dagger}, a_{s'}(x') \big\} = \delta_{xx'} \delta_{ss'}, 
  \quad 
  \big\{ a_{s}(x), a_{s'}(x') \big\} = 0, \quad \text{and} \quad
  \big\{ a_{s}(x)^{\dagger}, a_{s'}(x')^{\dagger} \big\} = 0, 
\end{equation*}
where $A^{\dagger}$ denotes the adjoint operator of $A$. Using the
second quantization formulation, the (many-body) Hamiltonian of the
Hubbard system is then given by
\begin{equation}\label{eq:HubbardHamiltonian}
  H = \sum_{x, y \in \ZZ^d} \sum_{s \in \{\uparrow, \downarrow\}} \alpha(x - y) a_{s}(x)^{\dagger} a_{s}(y) 
  + \tfrac{1}{2} \sum_{x \in \ZZ^d} \sum_{s, s' \in \{\uparrow, \downarrow\}} \lambda V(x-y) a_{s}(x)^{\dagger}a_{s}(x) a_{s'}(y)^{\dagger} a_{s'}(y).
\end{equation}
Here the first term on the right hand side in the Hamiltonian describes the hopping from site $y$ to $x$ with $\alpha$ the hopping amplitude, and the non-quadratic second term gives the interactions of two excitons with $V$ the interaction potential. The
grid $\ZZ^d$ must be distinguished from the spatial dimension
considered below (in some sense, the grid $\ZZ^d$ is on the
microscopic scale while the spatial inhomogeneity is introduced on a
mesoscopic scale). In Fourier representation, the time-dependent
(Heisenberg picture) field operators $\hat{a}_s(t,v)$ adhere to the
initial ($t = 0$) anti-commutation relation $\{
\hat{a}_s(v)^{\dagger}, \hat{a}_{s'}(v')\} = \delta_{s s'}\,\delta(v -
v')$, with $v$, $v'$ denoting velocity variables. As discussed in
\cite{DerivationBoltzmann2013}, the time-dependent average Wigner
matrix $W$ defined by
\begin{equation}\label{eq:defnW}
\big\langle \hat{a}_s(t,v)^{\dagger} \, \hat{a}_{s'}(t,v') \big\rangle = \delta(v - v') W_{s s'}(t,v)
\end{equation}
will approximately satisfy a Boltzmann kinetic equation $\partial_t W
= \mathcal{C}[W]$ for times up to order $\lambda^{-2}$. Here, as in
the Heisenberg picture, the average $\langle \cdot \rangle$ is taken
with respect to the initial state of the system. The effective
Boltzmann equation is much easier to solve compared to the original
quantum many-body system, which is an extremely high-dimensional
problem.


Augmenting the Boltzmann equation with the usual transport term for the spatially inhomogeneous setting and including an external magnetic field $\vec{B}$, one arrives at Eq.~\eqref{eq:boltzmann}, where $\vec{\sigma} =
(\sigma_1, \sigma_2, \sigma_3)$ are the Pauli matrices:
\begin{equation}
  \sigma_1 = 
  \begin{pmatrix} 0 & 1 \\ 1 & 0 \end{pmatrix}, \quad 
  \sigma_2 = 
  \begin{pmatrix} 0 & -i \\ i & 0 \end{pmatrix}, \quad 
  \sigma_3 = 
  \begin{pmatrix} 1 & 0 \\ 0 & -1 \end{pmatrix}. 
\end{equation}
Hence $\vec{B} \cdot \vec{\sigma} = \sum_{i=1}^3 B_i \sigma_i$.  In
\eqref{eq:boltzmann}, the Hermitian spin-density matrix Wigner
distribution $W: \RR_+ \times \Omega \times \RR^2 \to \CC^{2\times2}$
additionally depends on the spatial location $x \in \Omega$. When the
system is spatially homogeneous, $W$ as defined in \eqref{eq:defnW} is
a positive-semidefinite Hermitian matrix with eigenvalues in $[0, 1]$ at any $(t, v)$. In the case of
\eqref{eq:boltzmann} where spatial inhomogeneity is introduced on a
mesoscopic scale, we likewise assume that initially the Wigner matrix
$W(0, x, v)$ is positive-semidefinite with eigenvalues in $[0, 1]$
at any $(x, v)$. This property is preserved by the evolution of \eqref{eq:defnW}.

For simplicity, we will only consider the case that $\Omega$ is an
open and bounded interval on $\RR$; and assume that the velocity space
is $\RR^2$. In other words, we are considering the case with one space
dimension and two velocity dimensions. Physically, this means that the
solution is homogeneous with respect to one of the spatial variables
for a full two-dimensional (two space and two velocity dimensions)
model. Without loss of generality, we will also assume $\Omega =
(0,1)$.

The collision operator in \eqref{eq:boltzmann} consists of a conservative and dissipative
part: $\mathcal{C} = \mathcal{C}_{\mathrm{c}} + \mathcal{C}_{\mathrm{d}}$, where \cites{BoltzmannHubbard2012, BoltzmannNonintegrable2013}
\begin{equation}\label{eq:Cc}
  \mathcal{C}_{\mathrm{c}}[W](t, x, v) = -i \big[ H_{\eff}(t, x, v), W(t, x, v) \big]
\end{equation}
with the effective Hamiltonian
\begin{multline}\label{eq:Heff}
  H_{\eff}(t,x,v_1) = \int \ud v_2 \ud v_3 \ud v_4 \,
  \delta(\underline{v}) \mathcal{P}( 1 / \underline{\omega} ) (W_3 W_4 -
  W_2 W_3 - W_3 W_2 - \tr[W_4] W_3 + \tr[W_2] W_3 + W_2).
\end{multline}
Here $\mathcal{P}$ denotes the principal value, and we have used the
shorthand notations $W_i = W(t, x, v_i)$ for $i = 1, \dots, 4$,
the velocity difference $\underline{v} = v_1 + v_2 -
v_3 - v_4$, and the energy difference $\underline{\omega} =
\omega(v_1) + \omega(v_2) - \omega(v_3) - \omega(v_4)$.
The energy (or dispersion relation) $\omega(v)$ is precisely the Fourier transform of the hopping amplitude $\alpha$ appearing in \eqref{eq:HubbardHamiltonian}.
The terms $W_i W_j$ are usual matrix products.
Due to the invariance under $v_3 \leftrightarrow v_4$, the matrix product $W_3 W_4$ in the integrand could be replaced by $W_4 W_3$; in particular, $H_{\eff}$ is Hermitian.

The dissipative part of the collision operator is given by
\begin{equation}\label{eq:Cd}
  \mathcal{C}_{\mathrm{d}}[W](t, x, v_1) = \pi \int \ud v_2 \ud v_3 \ud v_4 \delta(\underline{v}) \delta(\underline{\omega}) \big( \mathcal{A}[W]_{1234} + \mathcal{A}[W]^{\ast}_{1234} \big)
\end{equation}
with
\begin{multline}
  \mathcal{A}[W]_{1234} = - W_4 \wt{W}_2 W_3 + W_4 \tr[\wt{W}_2 W_3] \\
  - \bigl(\wt{W}_4 W_3 - \wt{W}_4 W_2 - \wt{W}_2 W_3 +
  \wt{W}_4 \tr[W_2] - \wt{W}_4 \tr[W_3] + \tr[W_3
  \wt{W}_2]\bigr) W_1,
\end{multline}
where $\wt{W} = \mathbbm{1} - W$. As explained in \cites{BoltzmannHubbard2012, BoltzmannNonintegrable2013},
the first two summands (plus their Hermitian conjugates) can be identified as gain term and $( \dots ) W_1$ (plus Hermitian conjugate) as loss term.
Alternatively, by making use of the invariance under $v_3 \leftrightarrow v_4$, the integrand of $\mathcal{C}_{\mathrm{d}}[W]$ can be represented as
\begin{equation}
\mathcal{A}[W]_{1234} + \mathcal{A}[W]^{\ast}_{1234} \equiv \mathcal{A}_{\mathrm{quad}}[W]_{1234} + \mathcal{A}_{\mathrm{tr}}[W]_{1234}
\end{equation}
with
\begin{equation}
\label{eq:AW_Hubbard}
\begin{split}
\mathcal{A}_{\mathrm{quad}}[W]_{1234} &= -\tilde{W}_1 W_3 \tilde{W}_2 W_4 - W_4 \tilde{W}_2 W_3 \tilde{W}_1 + W_1 \tilde{W}_3 W_2 \tilde{W}_4 + \tilde{W}_4 W_2 \tilde{W}_3 W_1, \\
\mathcal{A}_{\mathrm{tr}}[W]_{1234} &= \big(\tilde{W}_1 W_3 + W_3 \tilde{W}_1\big) \tr[\tilde{W}_2 W_4] - \big(W_1 \tilde{W}_3 + \tilde{W}_3 W_1\big) \tr[W_2 \tilde{W}_4].
\end{split}
\end{equation}
The representation \eqref{eq:AW_Hubbard} emphasizes the similarity to the scalar collision operator \cite{UehlingUhlenbeck1933}, which is recovered when all $W_i$ are proportional to the identity matrix. For what follows, we always take $\omega(v) = \frac{1}{2} \abs{v}^2$ as dispersion relation. Note that if a general dispersion relation is taken, the transport term in \eqref{eq:boltzmann} changes to $(\partial_{v_x} \omega(v))\cdot \partial_x W$. 

One may check that the collision operator $\mathcal{C} = \mathcal{C}_{\mathrm{c}} +
\mathcal{C}_{\mathrm{d}}$ satisfies the density, momentum and energy conservation laws
(at each $x$)
\begin{equation}
  \int \mathcal{C}[W](v) \ud v = 0, \quad \int v \tr \bigl[\mathcal{C}[W](v)\bigr] \ud v = 0, \quad \int \tfrac{1}{2}\abs{v}^2 \tr\bigl[ \mathcal{C}[W](v) \bigr] \ud v = 0. 
\end{equation} 
As a result, the corresponding fluid dynamic moments, i.e., density $\rho(t, x) \in \CC^{2 \times 2}$, velocity $u(t, x) \in \RR^2$, and internal energy $\varepsilon(t, x) \in \RR$
\begin{align}
  \rho(t, x) & = \int W(t, x, v) \ud v; \label{eq:density}\\
  \tr[\rho(t, x)] u(t, x) & = \int v \tr [W(t, x, v)] \ud v; \label{eq:velocity}\\
  \tr[\rho(t, x)] \varepsilon(t, x) & = \int \tfrac{1}{2}\abs{v - u}^2 \tr[W(t, x, v)] \ud v \label{eq:internal_energy}
\end{align}
satisfy local conservation laws.

The (local) entropy of the state $W$ is defined as 
\begin{equation}\label{eq:entropy_def}
  S[W](t,x) = - \int \tr \bigl[ W(t,x,v) \log W(t,x,v) + \wt{W}(t,x,v) \log \wt{W}(t,x,v) \bigr] \ud v.
\end{equation}
The H-theorem states that the global entropy production rate is positive (see \cite{BoltzmannHubbard2012} for the matrix-valued case)
\begin{equation}
  \sigma[W](t) := \frac{\ud}{\ud t} \int S[W](t,x) \ud x = - \int \int \tr \Bigl[ \bigl(\log W(t,x,v) - \log \wt{W}(t,x,v) \bigr) \mathcal{C}[W](t,x,v) \Bigr] \ud v \ud x \geq 0
\end{equation}
for all $W$ with eigenvalues in $[0, 1]$ and periodic boundary conditions. The advection term in the integrand vanishes since we integrate over the spatial domain.

In the asymptotic long-time limit $t \to \infty$ for a closed system
(with periodic boundary conditions) and in the absence of
external fields, the solution of the Boltzmann equation
\eqref{eq:boltzmann} is expected to converge to the
Fermi-Dirac distribution
\begin{equation}\label{eq:W_FermiDirac}
  W_{\FD}(v) = \sum_{s \in \{ \uparrow, \downarrow \}} \bigl( \mathrm{e}^{( \omega(v) - \mu_{s})/(k_{\mathrm{B}}T)} + 1\bigr)^{-1} \ket{s}\bra{s}
\end{equation}
for a $v$-independent spin basis $\ket{s}$, temperature $T$, and
chemical potentials $\mu_{\uparrow}$ and $\mu_{\downarrow}$ (see
\cite{BoltzmannHubbard2012} for a proof of convergence in the
spatially homogeneous case). Note that the Fermi-Dirac distribution
maximizes the entropy among states with the same fluid dynamic
moments.  The moments of $W_{\FD}$ have analytical expressions: in two
dimensions and for the dispersion $\omega(v) = \frac{1}{2} \abs{v}^2$,
\begin{equation}\label{eq:momentsFermiDirac}
\rho_{\FD} = 2\pi k_{\mathrm{B}} T \sum_{s \in \{ \uparrow, \downarrow \}} \log\bigl(1 + \mathrm{e}^{\mu_s / (k_{\mathrm{B}}T)} \bigr), \quad \varepsilon_{\FD} = k_{\mathrm{B}} T  \frac{-\sum_{s \in \{ \uparrow, \downarrow \}} \mathrm{Li}_2\bigl(-\mathrm{e}^{\mu_s / (k_{\mathrm{B}}T)} \bigr)}{\sum_{s \in \{ \uparrow, \downarrow \}} \log\bigl(1 + \mathrm{e}^{\mu_s / (k_{\mathrm{B}}T)} \bigr)}
\end{equation}
where $\mathrm{Li}_n$ is the polylogarithm function. The average velocity of $W_{\FD}$ in \eqref{eq:W_FermiDirac} is zero.

To complete the equation, we need to impose the boundary
conditions. Let $\Sigma = \partial \Omega \times \RR^2 = \{0, 1\}
\times \RR^2$, and denote by $n(x)$ be the outward unit normal vector at $x
\in \partial \Omega$. We define the outgoing and incoming boundaries
as
\begin{equation}
  \Sigma_{\pm} = \big\{ (x,v) \in \Sigma; \pm n(x) \cdot v > 0 \big\}.
\end{equation}
Hence 
\begin{align}
  & \Sigma_+ = \big\{ (0,v); v_x < 0\} \cup \{(1, v); v_x > 0 \big\}; \\
  & \Sigma_- = \big\{ (0,v); v_x > 0\} \cup \{(1, v); v_x < 0 \big\}.
\end{align}
For the boundary condition on the incoming boundary $\Sigma_-$, we consider
\begin{itemize}
\item periodic boundary conditions: for $(x, v) \in \Sigma_-$,
  \begin{equation}
    W(t, x, v) = W(t, 1-x, v).
  \end{equation}
  Note that $(1-x, v) \in \Sigma_+$. 
\item Dirichlet boundary conditions: for $(x, v) \in \Sigma_-$, 
 \begin{equation}
    W(t, x, v) = \Phi(t, x, v)
  \end{equation}
  where $\Phi: \RR_+ \times \Sigma_- \to \CC^{2\times2}$ is a given
  boundary state.
\item Maxwell boundary conditions:
  \begin{equation}
    W(t, x, v) = \mathcal{R}_x\big( W(t, x, \cdot)\vert_{\Sigma_+^x}\big)(v),
  \end{equation}
  where $\mathcal{R}_x$ is a Maxwell reflection operator:
  \begin{equation}
    \mathcal{R}_x = (1 - \alpha) \mathcal{L}_x + \alpha \mathcal{D}_x. 
  \end{equation}
  Here $\alpha \in [0,1]$ is the accommodation coefficient. The local
  reflection operator $\mathcal{L}_x$ is given by 
  \begin{equation}
    \big(\mathcal{L}_x F\big)(v_x, v_y) = F(-v_x, v_y), 
  \end{equation}
  and the diffusive reflection is given by (for a specified
  Fermi-Dirac state depending on spin basis, $T$, $\mu_{\uparrow}$
  and $\mu_{\downarrow}$)
  \begin{equation}
    (\mathcal{D}_x F)(v) = Z_x^{-1} W_{\FD}(v) \wb{F}(x),
  \end{equation}
  where $\wb{F}(x)$ is the total outgoing number flux 
  \begin{equation}
    \wb{F}(x) = \int_{v \cdot n(x)>0} \tr(F(v)) v\cdot n(x) \ud v,
  \end{equation}
  and $Z_x$ is a normalizing constant such that 
  \begin{equation}
    Z_x = \int_{v \cdot n(x)<0} \tr(W_{\FD}(v)) \abs{v \cdot n(x)} \ud v. 
  \end{equation}
\end{itemize}
In the following, we introduce an efficient and accurate numerical
scheme for the Boltzmann equation \eqref{eq:boltzmann}.

\section{Fast spectral method for the collision operator}
\label{sec:collision}

Calculating the collision operator is the computationally most demanding step in solving \eqref{eq:boltzmann}. Here, we first represent the collision operator using Pauli matrices, and then develop a fast Fourier spectral method inspired by the ideas in \cites{MouhotPareschi:06, FilbertHuJin:12, HuYing:12}.

\subsection{Representation of the collision operator using Pauli matrices}

Since the spin-density matrix $W(t, x, v)$ is Hermitian, it can be
represented in the basis of the identity matrix and the Pauli matrices:
\begin{equation}
\label{eq:spin_repr}
W_i = W(v_i) = w_{i, 0} \mathbbm{1} + \sum_{j = 1}^3 w_{i, j} \sigma_{j},
\end{equation}
where the subscript $i$ specifies the velocity dependence, and we have
suppressed the dependence on $(t, x)$ for concise notation.  Moreover,
we define the vector of components as
\begin{equation}
w_i = (w_{i,0}, w_{i,1}, w_{i,2}, w_{i,3}) \in \mathbb{R}^4, 
\end{equation}
and introduce the notation $\bd{\sigma} = (\mathbbm{1},
\vec{\sigma})$ so that 
\begin{equation}
  w_i \cdot \bd{\sigma} = w_{i, 0} \mathbbm{1} + \sum_{j=1}^3 w_{i, j} \sigma_j.
\end{equation}
The $3$-vector part $\vec{w}_i = (w_{i,1}, w_{i,2}, w_{i,3}) \in \mathbb{R}^3$
is exactly the Bloch vector of $W_i$ (up to normalization), and the eigenvalues of $W_i$ are $w_{i,0} \pm \abs{\vec{w}_i}$.

We will also use the $4 \times 4$ ``metric tensor'' $\eta = \mathrm{diag}(1, -1, -1, -1)$ and set
\begin{equation}
\langle w_i, w_j \rangle_{\eta} = w_i^T \eta\, w_j.
\end{equation}
Since the eigenvalues of $W_i$ are in the interval $[0,1]$, one can verify that likewise $\langle w_i, w_j \rangle_{\eta} \in [0,1]$.

Using the interchangeability of $v_3 \leftrightarrow v_4$ in the integral,
the gain term of the dissipative collision operator \eqref{eq:Cd} can be written as
\begin{equation}
- W_4 \wt{W}_2 W_3 + W_4 \tr[\wt{W}_2 W_3] + \mathrm{h.c.}
\equiv 2\, \langle w_3, w_4 \rangle_{\eta} \big( \mathbbm{1} - (\eta\, w_2)\cdot\bd{\sigma} \big),
\end{equation}
and the loss term as
\begin{multline}
- \bigl(\wt{W}_4 W_3 - \wt{W}_4 W_2 - \wt{W}_2 W_3
+ \wt{W}_4 \tr[W_2] - \wt{W}_4 \tr[W_3] + \tr[W_3 \wt{W}_2]\bigr) W_1 + \mathrm{h.c.}\\
\equiv - \big( \langle w_3, w_4 \rangle_{\eta} \mathbbm{1}
- (w_{3,0} + w_{4,0} - 1) \, (\eta\, w_2)\cdot\bd{\sigma} \big) W_1 + \mathrm{h.c.}
\end{multline}
For the Hermitian conjugate, one requires the
anti-commutator of two Wigner matrices, which reads in the Pauli
representation
\begin{equation}
  \bigl\{ W_1, W_2 \bigr\} \equiv W_1 W_2 + W_2 W_1 = 2\, \big( (w_{1,0} w_2 + w_{2,0} w_1) \cdot \bd{\sigma} - \langle w_1, w_2 \rangle_{\eta} \mathbbm{1} \big).
\end{equation}

Again using the interchangeability of $v_3 \leftrightarrow v_4$ in the integral, the integrand in Eq.~\eqref{eq:Heff} of the conservative collision operator \eqref{eq:Cc} becomes
\begin{multline}
(W_3 W_4 - W_2 W_3 - W_3 W_2 - \tr[W_4] W_3 + \tr[W_2] W_3 + W_2) \\
\equiv \big( \langle w_2, w_2 \rangle_\eta - \langle w_3 - w_2, w_4 - w_2 \rangle_\eta \big) \mathbbm{1} - (w_{3,0} + w_{4,0} - 1) W_2.
\end{multline}
Since the identity matrix does not contribute to the commutator
in Eq.~\eqref{eq:Cc}, it suffices to keep the second term $-
(w_{3,0} + w_{4,0} - 1) W_2$ only. The commutator in Eq.~\eqref{eq:Cc}
reads in the Pauli matrix representation
\begin{equation}
-i\,\bigl[W_2, W_1\bigr] = 2\, (\vec{w}_2 \times \vec{w}_1) \cdot \vec{\sigma}.
\end{equation}

To summarize, we have obtained the representation
\begin{equation}\label{eq:CcPauli}
  \mathcal{C}_{\mathrm{c}}[W]_1 = i \int \ud v_2 \ud v_3 \ud v_4 \, \delta(\underline{v}) \mathcal{P}( 1 / \underline{\omega} ) (w_{3, 0} + w_{4, 0} - 1) \bigl[W_2, W_1\bigr],
\end{equation}
and similarly
\begin{multline}\label{eq:CdPauli}
  \mathcal{C}_{\mathrm{d}}[W]_1 = \pi \int \ud v_2 \ud v_3 \ud v_4 \, 
  \delta(\underline{v}) \delta(\underline{\omega})
  \Bigl( 2 \average{w_3, w_4}_{\eta} \bigl(\mathbbm{1} - W_1 - (\eta w_2) \cdot \bd{\sigma}\bigr) \\
  + (w_{3,0} + w_{4,0} - 1) \bigl\{ W_1, ( \eta w_2) \cdot \bd{\sigma}
  \bigr\} \Bigr).
\end{multline}

We remark that it is also possible to write the matrix-valued
Boltzmann equation as a kinetic equation with multiple components, if
we regard each matrix entry as a component. However, we prefer the
more natural and physical representation in terms of Pauli matrices.

\subsection{Fast Fourier spectral method}
\label{sec:FourierCollision}

To efficiently evaluate the collision terms \eqref{eq:CcPauli} and
\eqref{eq:CdPauli}, we generalize the ideas in
\cites{MouhotPareschi:06, FilbertHuJin:12, HuYing:12} for a Fourier
spectral discretization of the velocity space.

Let us discuss the conservative part \eqref{eq:CcPauli} first. Using
the Carleman representation \cites{Carleman:57, Wennberg:94}, we perform a change of variables $v_1 \mapsto v,  v_3 \mapsto v + u, v_4 \mapsto v + u'$, such that
\begin{align*}
  & v_2 =  v_3 + v_4 - v_1 = v + u + u'; \\
  & \underline{\omega} = \omega(v_1) + \omega(v_2) - \omega(v_3) - \omega(v_4)
  = u \cdot u'. 
\end{align*}
Substituting into \eqref{eq:CcPauli}, we arrive at 
\begin{equation}\label{eq:CcPauliCarleman}
  \mathcal{C}_{\mathrm{c}}[W](v) = i \int_{B_R} \int_{B_R} \ud u \ud u' \mathcal{P}( 1 / (u \cdot u')) (2 w_0(v+u)  - 1) \bigl(W(v+u+u')W(v) - \text{h.c.}\bigr),
\end{equation}
where we have used the symmetry between $u$ and $u'$. Here $R$ indicates the
truncation of the collision integral, taken so that $B_R$
approximately covers the support of $W$ in the $v$ variable.  Hence, we just need
to deal with integrals of the kind
\begin{equation}
  I_1(v) = \int_{B_R} \int_{B_R} \ud u \ud u' \mathcal{P}(1 / (u \cdot u')) f(v+u) g(v + u + u') h(v). 
\end{equation}
To apply the Fourier method, we periodize the functions $f$, $g$ and
$h$ \text{etc.} on the domain $[-L, L]^2$ with $L \geq \frac{3 +
  \sqrt{2}}{2} R$, and define the Fourier grid
\begin{equation}\label{eq:fourier_grid}
  \Xi = [-N/2, -N/2 + 1, \ldots, N/2 - 1]^2.
\end{equation}
Here the cut-off frequency $N$ controls the accuracy.
Using the Fourier inversion formula, we approximate
\begin{equation}
  f(v) \approx \sum_{\xi \in \Xi} \wh{f}(\xi) \exp( i \pi \xi \cdot v / L)
  \quad \text{and} \quad \wh{f}(\xi) = \slashint_{[-L, L]^2} f(v) \mathrm{e}^{-  i \frac{\pi}{L} \xi \cdot v} \ud v
\end{equation}
Then 
\begin{equation}
  \begin{split}
    I_1(v) & = \sum_{\chi, \eta, \zeta} \int_{B_R} \int_{B_R} \ud u \ud u' \mathcal{P}(1 / (u
    \cdot u')) \wh{f}(\chi) \wh{g}(\eta) \wh{h}(\zeta) \mathrm{e}^{i
      \frac{\pi}{L} v \cdot (\chi + \eta + \zeta)}
    \mathrm{e}^{i \frac{\pi}{L} u \cdot ( \chi + \eta)} \mathrm{e}^{i \frac{\pi}{L} u' \cdot \eta} \\
    & = \sum_{\chi, \eta, \zeta} \wh{f}(\chi) \wh{g}(\eta) \wh{h}(\zeta) \mathrm{e}^{i
      \frac{\pi}{L} v \cdot (\chi + \eta + \zeta)}
    G(\chi+\eta, \eta) 
  \end{split}
\end{equation}
where the matrix $G(\xi, \chi)$ is defined as 
\begin{equation}
  G(\xi, \chi) =  \int_{B_R} \ud u \int_{B_R} \ud u'\, \mathcal{P}(1 / (u\cdot u')) \exp(i \pi \xi \cdot u / L)  \exp(i \pi \chi \cdot u' / L). 
\end{equation}
Changing to polar coordinates, one obtains
\begin{equation}
  G(\xi, \chi) = \int_0^R \ud r \int_0^R \ud r' \int_{S^1} \ud \theta \int_{S^1} \ud \theta' \, \mathcal{P}\bigl(1 / (\theta \cdot \theta')\bigr) \exp\bigl(i \pi \xi \cdot \theta r/ L \bigr)  \exp\bigl(i \pi \chi \cdot \theta' r'/ L \bigr). 
\end{equation}
Since $\mathcal{P}(1/(\theta \cdot \theta'))$ is odd in both $\theta$ and $\theta'$, it suffices to take the odd part of the complex exponentials in the above integral, and we get 
\begin{equation}
  G(\xi, \chi) = - \int_0^R \ud r \int_0^R \ud r' \int_{S^1} \ud \theta \int_{S^1} \ud \theta' \, \mathcal{P}\bigl(1 / (\theta \cdot \theta')\bigr) \sin\bigl(\pi \xi \cdot \theta r/ L \bigr)  \sin\bigl(\pi \chi \cdot \theta' r' / L \bigr). 
\end{equation}
Note that 
\begin{equation}
  \phi_R(\xi \cdot \theta) = \int_0^R \ud r \sin\bigl(\pi \xi \cdot \theta r / L \bigr) 
  = \frac{L}{\pi (\xi \cdot \theta)} \Bigl[ \cos\bigl(\pi \xi \cdot \theta R / L \bigr) - 1 \Bigr]  
  = - \frac{2L}{\pi(\xi \cdot \theta)} \sin^2\bigl(\pi \xi \cdot \theta R / (2L)\bigr).
\end{equation}
Hence, 
\begin{equation}
    G(\xi, \chi) = -  \int_{S^1} \ud \theta \int_{S^1} \ud \theta' \mathcal{P}\bigl(1 / (\theta \cdot \theta')\bigr) \phi_R(\xi \cdot \theta) \phi_R(\chi \cdot \theta'). 
\end{equation}
We approximate $G$ using numerical quadrature with a trapezoidal rule. To deal with the singularity in the principal value integral, we take the grids of $\theta$ and $\theta'$ to be 
\begin{equation}\label{eq:theta_grid}
  \theta_j = \exp(i(j-1) \pi / J), \quad \text{and} \quad 
  \theta_j' = \exp(i(j-1/2) \pi / J), \qquad j = 1, \ldots, J
\end{equation}
for some positive integer $J$. The quadrature rule converges
exponentially \cite{TrefethenWeideman:13}*{Section 6}. Due to
symmetry, only quadrature points on half circles are
required. We arrive at the final approximation
\begin{equation}\label{eq:approxG}
  G(\xi, \chi) \approx \sum_{j = 1}^{J} \sum_{j' = 1}^J \omega_{G, j, j'}    \phi_{R, j}(\xi)   \phi'_{R, j'}(\chi)
\end{equation}
where the weights are given by 
\begin{equation}
  \omega_{G, j, j'} = - \Bigl( \frac{2 \pi}{J}\Bigr)^2 \frac{1}{\theta_j \cdot \theta'_{j'}}
\end{equation}
and we have used the shorthand notation
\begin{equation}
  \phi_{R, j}(\xi) = \phi_R(\xi \cdot \theta_j) \quad \text{and} \quad
  \phi'_{R, j'}(\chi) = \phi_R(\chi \cdot \theta'_{j'}).
\end{equation}
In summary, we have obtained the approximation 
\begin{equation}
  \begin{split}
    \wh{I}_1(\xi) & \approx \sum_{\substack{\chi, \eta, \zeta, \\ \chi + \eta + \zeta = \xi}} \sum_{j, j'} \omega_{G, j, j'}\, \wh{f}(\chi)\,\wh{g}(\eta)\,\wh{h}(\zeta)\,\phi_{R, j}(\chi + \eta)\, \phi'_{R, j'}(\eta) \\
    & = \sum_{j, j'} \omega_{G, j, j'} \sum_{\zeta} \Bigl[ \sum_{\eta}
    \wh{f}(\xi - \zeta - \eta)\, \bigl(\phi'_{R, j'}(\eta)\, \wh{g}(\eta)
    \bigr) \Bigr] \phi_{R, j}(\xi - \zeta) \, \wh{h}(\zeta).
  \end{split} 
\end{equation}
For each $j$ and $j'$, we first calculate the product $\phi'_{R, j'}
\wh{g}$ (complexity $\Or(JN^2)$); the summation over $\eta$ is a convolution by
FFT ($\Or(JN^2 \log N)$); we then multiply the result pointwise with $\phi_{R,
  j}$ ($\Or(J^2N^2)$). The summation over $\zeta$ is another
convolution ($\Or(J^2N^2 \log N)$). The total complexity is thus $\Or(J^2
N^2 \log N)$. We use double zero padding in the Fourier
coefficients to avoid aliasing.

\smallskip 

The dissipative part \eqref{eq:CdPauli} in Carleman
representation reads 
\begin{multline}
  \mathcal{C}_{\mathrm{d}}[W](v) = \pi \int_{B_R} \int_{B_R} \ud u \ud u' \, 
  \delta(u \cdot u')
  \Bigl( 2 \average{w(v+u), w(v+u')}_{\eta} \bigl(\mathbbm{1} - W(v) - (\eta w(v+u+u')) \cdot \bd{\sigma}\bigr) \\
  + (2 w_{0}(u+v) - 1) \bigl\{ W(v), ( \eta w(v+u+u')) \cdot \bd{\sigma}
  \bigr\} \Bigr), 
\end{multline}
which follows from \eqref{eq:CdPauli} by the same change of variables leading to \eqref{eq:CcPauliCarleman}. 
Expanding the above expression, it is straightforward to check that it
consists of the following three kinds of integrals:
\begin{align}
  & I_2(v) = \int_{B_R} \int_{B_R}  \ud u \ud u' \delta(u \cdot u') f(v+u) g(v+u') h(v); \\
  & I_3(v) = \int_{B_R} \int_{B_R} \ud u \ud u' \delta(u \cdot u') f(v+u) g(v+u') h(v + u + u'); \\
  & I_4(v) = \int_{B_R} \int_{B_R} \ud u \ud u' \delta(u \cdot u')
  f(v+u) g(v+u + u') h(v),
\end{align}
where $f, g, h$ stand for certain components of $w$.  To evaluate these
integrals, we define
\begin{equation}
  H(\xi, \chi) = \int_{B_R} \ud u \int_{B_R} \ud u'\, \delta(u\cdot u') \exp(i \pi \xi \cdot u / L)  \exp(i \pi \chi \cdot u' / L). 
\end{equation}
By similar steps as leading to \eqref{eq:approxG}, one obtains
\begin{equation}\label{eq:approxH-J}
  H(\xi, \chi) \approx \sum_{j = 1}^{J} \omega_{H, j}\, \psi_{R, j}(\xi)\, \psi'_{R, j}(\chi). 
\end{equation}
Here
\begin{equation}
  \omega_{H, j} = \frac{\pi}{J}, \quad \psi_{R, j}(\xi) = \frac{2L}{\pi \xi \cdot \theta_j} \sin\Bigl( \frac{\pi R}{L} \xi \cdot \theta_j \Bigr) \quad \text{and} \quad \psi'_{R, j}(\chi) = \frac{2L}{\pi \xi \cdot \theta_j} \sin\Bigl( \frac{\pi R}{L} \xi \cdot \mathcal{R}_{\pi/2}\theta_j \Bigr),
\end{equation}
where $\mathcal{R}_{\pi/2}$ is the rotation by $\pi/2$. 

Then the Fourier representation of $I_2$ is
\begin{equation}
  \begin{split}
    \wh{I}_2(\xi) & = \sum_{\substack{\chi, \eta, \zeta, \\ \chi + \eta + \zeta = \xi}} \sum_{j} \omega_{G,j} \wh{f}(\chi)\, \wh{g}(\eta)\, \wh{h}(\zeta)\, \psi_{R, j}(\chi)\, \wt{\psi}_{R, j}(\eta) \\
    & = \sum_j \omega_{G, j} \sum_{\zeta} \sum_{\chi + \eta = \xi - \zeta} \bigl(\psi_{R, j}(\chi)\, \wh{f}(\chi) \bigr) \bigl( \wt{\psi}_{R, j}(\eta)\, \wh{g}(\eta) \bigr)\, \wh{h}(\zeta).
  \end{split}
\end{equation}
For each $j$, the pointwise products in the brackets are computed (complexity $\Or(JN^2)$); for each $j$ and $\xi$, summation over $\chi$, $\eta$ and $\zeta$ is a double convolution by FFT $(\Or(JN^2 \log N))$. The total complexity is $(\Or(J N^2 \log N))$. The integral type $I_4$ is similar to $I_1$, and hence we will omit the details. 

\smallskip 

For $I_3$, we need another representation of $H(\xi, \chi)$, as suggested by Hu and Ying in \cite{HuYing:12}:
\begin{equation}\label{eq:approxH-JM}
  \begin{aligned}
    H(\xi, \chi) & \approx \frac{\pi}{J} \sum_{m=1}^M w_{R, m}
    \sum_{j=1}^J \Bigl[ \exp( i \pi \rho_{R, m} \xi \cdot \theta_j  / L)\, \psi_{R, j}'(\chi) \Bigr], \\
    & =:  \sum_{m=1}^M \sum_{j=1}^J \wt{\omega}_{H, j, m}
     \Bigl[ \exp( i \alpha_{j, m} \cdot \xi)\, \psi_{R, j}'(\chi) \Bigr]
  \end{aligned}
\end{equation}
with 
\begin{equation}
  \wt{\omega}_{H, j, m} = \frac{\pi}{J} w_{R, m} \quad \text{and} \quad 
  \alpha_{j, m} = \pi \rho_{R, m} \theta_j / L. 
\end{equation}
In the above, $(\rho_{R, m}, w_{R, m})$ are the nodes and weights of a
Gauss-Legendre quadrature on $[-R, R]$.  Then, the integral $I_3$ can
be calculated as
\begin{equation}
  \begin{aligned}
    \wh{I}_3(\xi) & = \sum_{j, m} \wt{\omega}_{H, j, m} \sum_{\chi,
      \eta, \zeta: \, \chi + \eta + \zeta = \xi} \wh{f}(\chi)\,
    \wh{g}(\eta)\, \wh{h}(\zeta)\, \exp(i \alpha_{j, m} \cdot \chi)\,
    \exp(i \alpha_{j, m} \cdot \zeta)\, \psi'_{R, j}(\eta + \zeta) \\
    & = \sum_{j, m} \wt{\omega}_{H, j, m} \sum_{\chi, \eta, \zeta: \,
      \chi + \eta + \zeta = \xi} \bigl(\exp(i \alpha_{j, m} \cdot
    \chi) \wh{f}(\chi)\bigr)\,
    \wh{g}(\eta)\, \psi'_{R, j}(\xi - \chi)\, \bigl(\exp(i \alpha_{j, m} \cdot \zeta)\, \wh{h}(\zeta) \bigr)  \\
    & = \sum_{j, m} \wt{\omega}_{H, j, m} \sum_{\chi} \bigl(\exp(i
    \alpha_{j, m} \cdot \chi)\, \wh{f}(\chi)\bigr)\, \psi'_{R, j}(\xi -
    \chi) \sum_{\eta, \zeta: \, \eta + \zeta = \xi - \chi} \wh{g}(\eta)\,
    \bigl(\exp(i \alpha_{j, m} \cdot \zeta)\, \wh{h}(\zeta) \bigr).
  \end{aligned}
\end{equation}
We first calculate the pointwise products in the bracket for each $j$ and $m$ ($\Or(JMN^2)$); summation over $\chi$, $\eta$  and $\zeta$ is a double convolution $(\Or(JMN^2 \log N))$. Since $M$ is $\Or(N)$, the total complexity is $(\Or(JN^3 \log N))$.

\smallskip

We remark that while the above discussion is limited to the case of
two dimensional velocity space, it is possible to extend the method to
$3D$ by generalizing the method in \cite{HuYing:12} to our case. We
will leave this to future works. The numerical results in this work
are limited to one space dimension and two velocity dimensions.

\section{Time splitting algorithm and parallelization}
\label{sec:parallel}

The numerical discretization of Eq.~\eqref{eq:boltzmann} is based on a
time splitting algorithm to deal with the convection, collision and
external magnetic terms separately.  Specifically, we perform half a
time step of convection in physical velocity space, then transform the
Wigner state to velocity Fourier space using FFT for the collision
integrals as discussed in Section~\ref{sec:FourierCollision} and for
applying the external magnetic field, and finally switch back to
physical velocity space for another half time step of convection. For
the convection we use the finite volume method with minmod slope
limiter \cite{LeVeque:92}*{Ch.~16}; other slope limiters can also be
applied. For completeness, we briefly recall the formulation with some
comments on parallelization of the algorithm.

In the advection step, each discrete velocity $v$ can be treated independently due to the time splitting. The minmod slope limiter method updates the solution as
\begin{equation}
\label{eq:slopeLimiterStep}
\begin{split}
W_j^{n+1} &= W_j^n + \frac{\Delta t}{2\Delta x} \Bigg[ -v_x \big(W_{j+1}^n - W_{j-1}^n\big) + \abs{v_x}\big(W_{j+1}^n - 2 W_j^n + W_{j-1}^n \big) \\
&\hspace{8em} + \tfrac{1}{2}\left(\mathrm{sgn}(\hat{\nu}) - \hat{\nu}\right) \left( -v_x \big(S_{j+1}^n - S_{j-1}^n \big) + \abs{v_x}\big( S_{j+1}^n - 2 S_j^n + S_{j-1}^n \big) \right) \Bigg].
\end{split}
\end{equation}
where $\Delta x$ denotes the spatial mesh width, $\Delta t$ the time
step, $n$ the discretized time index, $j$ the finite volume
cell index, and $\hat{\nu} = \Delta t\, v_x / \Delta x$. We use the shorthand notations
\begin{equation*}
W_j^n = W(n\,\Delta t, j\,\Delta x, v) \quad \text{and} \quad 
S_j^n = \minmod\big(W_{j+1}^n - W_j^n, W_j^n - W_{j-1}^n\big)
\end{equation*}
with the minmod function defined as
\begin{equation}
\label{eq:minmod}
\minmod(a,b) = \begin{cases}
a & \mathrm{if}\ \ \abs{a} \le \abs{b} ~~\text{and}~~ a\,b > 0; \\
b & \mathrm{if}\ \ \abs{b} < \abs{a} ~~\text{and}~~ a\,b > 0; \\
0 & \mathrm{if}\ \ a\,b < 0.
\end{cases}
\end{equation}

The first line in Eq.~\eqref{eq:slopeLimiterStep} is precisely Godunov's method, and the second line originates from the additional slope limiter terms. As mentioned above, for the time splitting algorithm we actually perform two transport steps with $\Delta t/2$. Note that the Wigner state at the next time step depends on \emph{two} neighbors on either side, i.e., on the five finite volumes with indices $j-2, \dots, j+2$.

Concerning parallelization, each computing node handles a few adjacent finite volumes, that is, we parallelize the computation along the spatial $x$ dimension. This straightforward approach takes advantage of the locality of the computationally demanding collision step, which is independent of the neighboring finite volumes. Since the transport step \eqref{eq:slopeLimiterStep} depends on two neighbors on each side, every computing node handles at least two adjacent finite volumes to minimize inter-process communication. In our custom C implementation, we use MPI to transfer neighboring states during the transport step.

\section{Numerical examples}\label{sec:numerics}

\subsection{Validation of the algorithm} 

We first present several numerical tests to validate our algorithm. Let us start with the approximation of the kernels $G$ in \eqref{eq:approxG} and $H$ in \eqref{eq:approxH-J} and \eqref{eq:approxH-JM}, which depend on the choice of the number $2 J$ of points on the unit circle in \eqref{eq:theta_grid}, and the total number $M$ of Gauss-Legendre quadrature nodes on $[-R, R]$ for the radial quadrature in \eqref{eq:approxH-JM}. To test the dependence on these parameters, we calculate the dissipative and conservative collision kernels $\mathcal{C}_{\mathrm{d}}[W]$ and $\mathcal{C}_{\mathrm{c}}[W]$ for a fixed spin density matrix $W(v)$. Here, $W(v)$ is (somewhat arbitrarily) chosen as
\begin{multline}\label{eq:validationW}
  W(v) =
  \frac{1}{3\pi} (1 + v_x) \mathrm{e}^{-\tfrac{1}{2}(v_x - 1/2)^2 - \tfrac{1}{2}(v_y + v_x)^2} \mathbbm{1}
  + \frac{3\sqrt{6/5}}{55\pi} \left(1 - v_x - \tfrac{v_y}{6}\right)^2 \mathrm{e}^{-\tfrac{1}{12}(2 v_x + v_y)^2 - \tfrac{1}{10} v_y^2} \sigma_1 \\
  + \frac{1}{4\pi \sqrt{2}} \mathrm{e}^{-\tfrac{1}{8} (v_x-1)^2 - \tfrac{1}{4} v_y^2} \sigma_2
  + \frac{9}{56\pi} \left(\tfrac{1}{3} + v_x\right)^2 \mathrm{e}^{-\tfrac{1}{8} (v_x-v_y)^2 - \tfrac{1}{8} (v_x + 2 v_y)^2} \sigma_3 \ .
\end{multline}
The Fourier representation determines the discretization in physical velocity space, i.e., $v \in (2 L/N)\,\Xi$ with $N$ the number of Fourier grid points in each dimension, $\Xi$ the corresponding grid defined in \eqref{eq:fourier_grid} and $L$ the domain size.

\begin{figure}[!ht]
\centering
\subfloat[convergence with $J$]{
\label{fig:conv-J}
\includegraphics[width=0.3\textwidth]{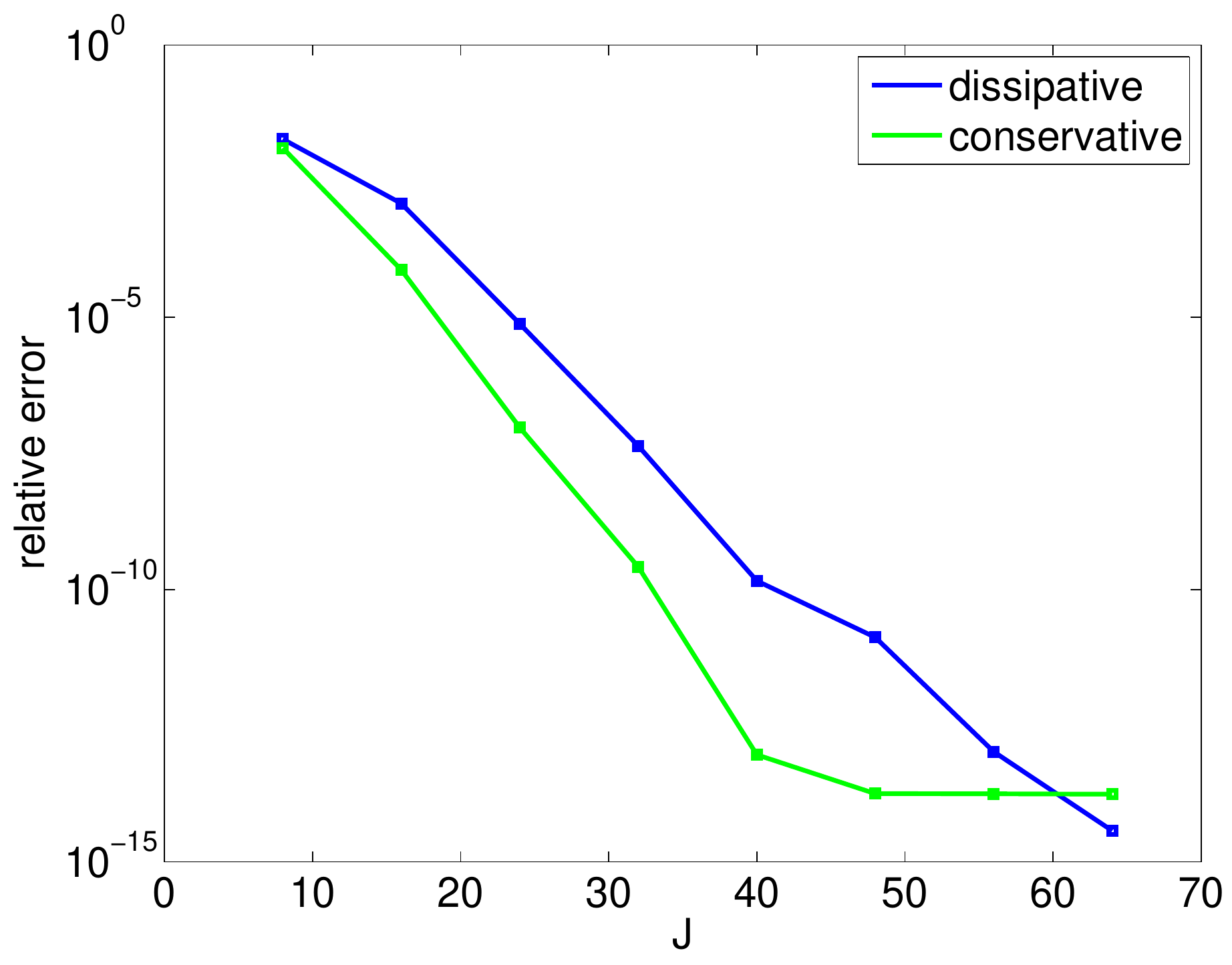}}
\hspace{0.03\textwidth}
\subfloat[convergence with $M$]{
\label{fig:conv-M}
\includegraphics[width=0.3\textwidth]{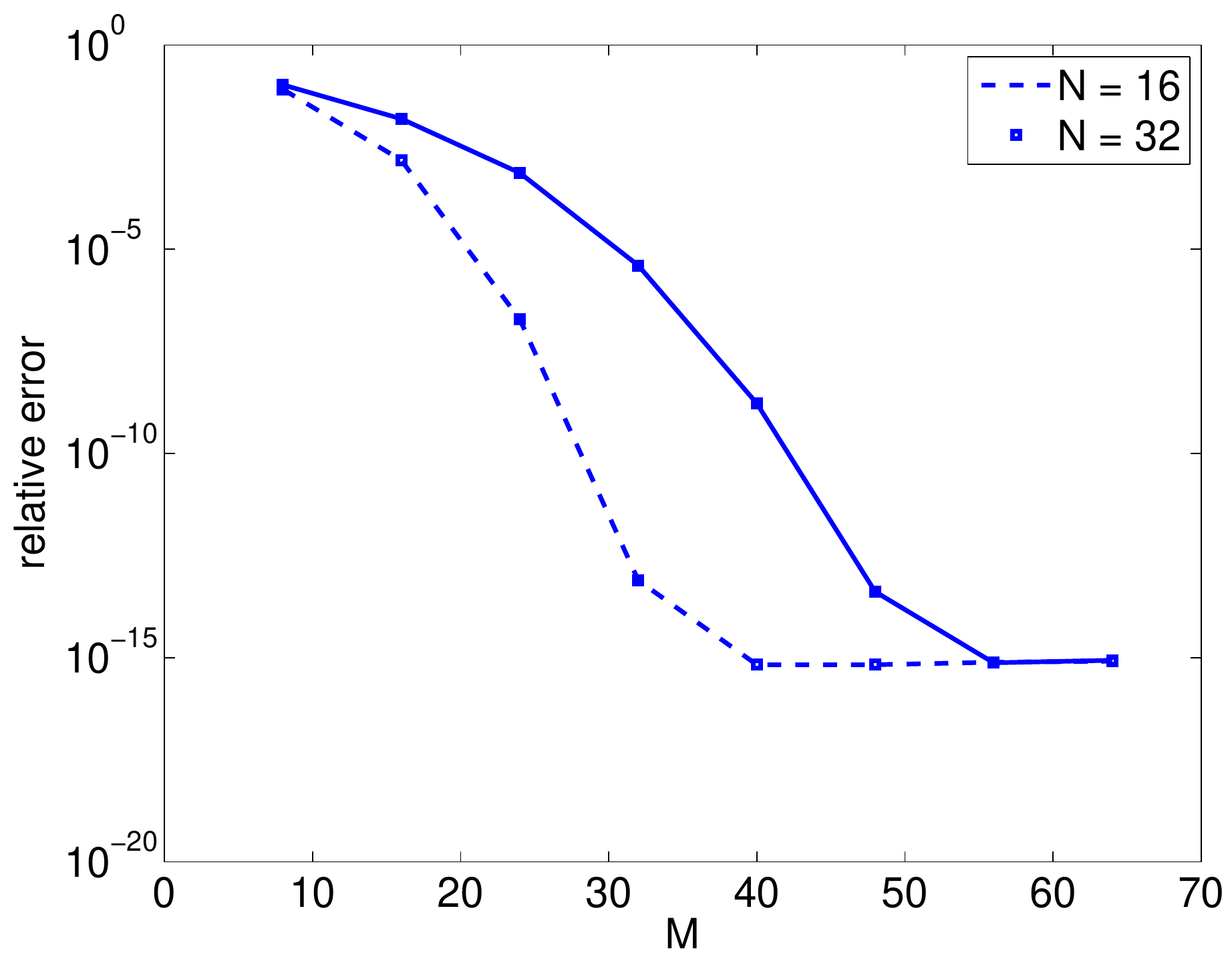}}
\hspace{0.03\textwidth}
\subfloat[convergence with $N$]{
\label{fig:conv-N}
\includegraphics[width=0.3\textwidth]{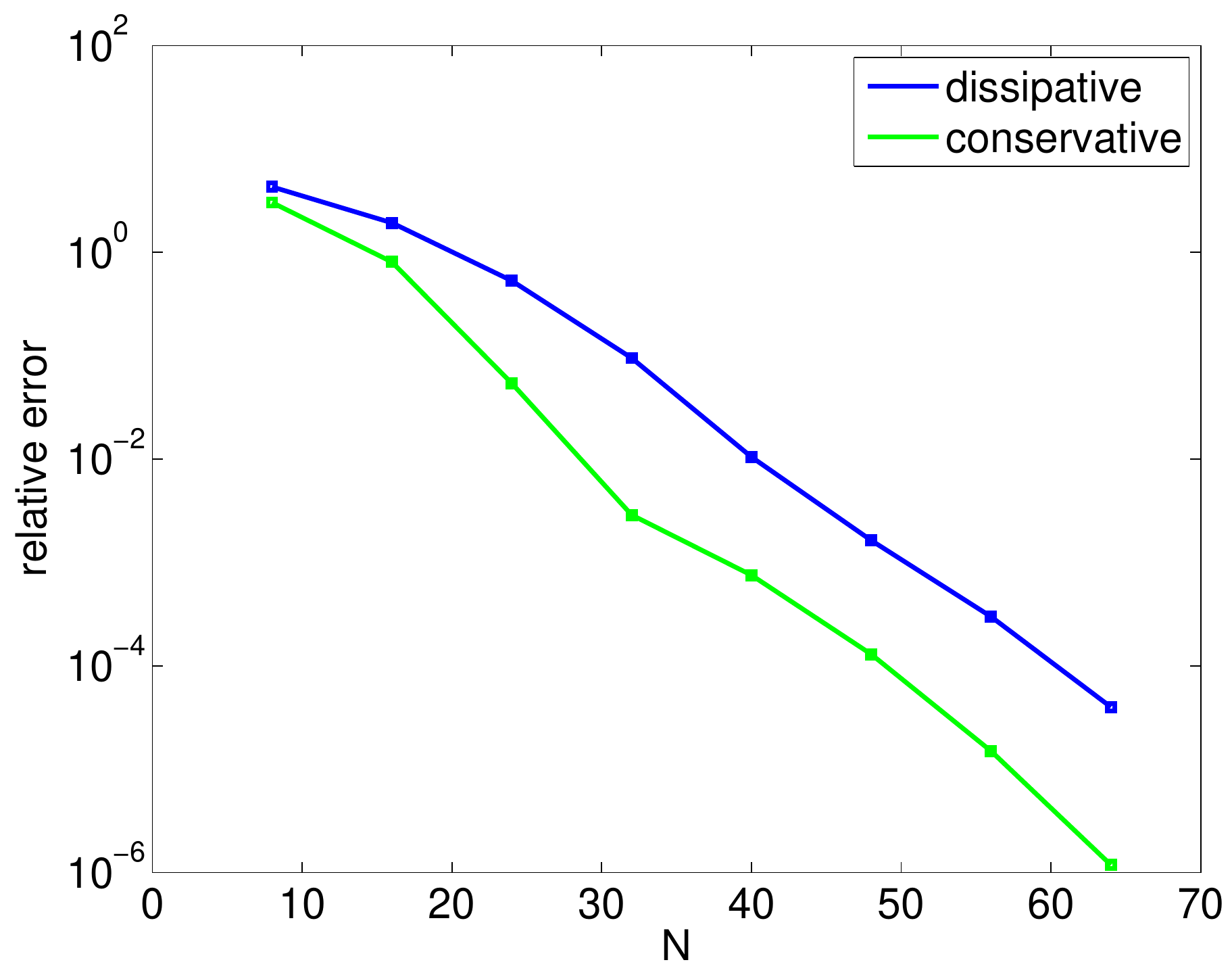}}
\caption{Exponential convergence of the dissipative $\mathcal{C}_{\mathrm{d}}$ and conservative $\mathcal{C}_{\mathrm{c}}$ collision operator calculation with respect to $J$, $M$ and $N$, for fixed domain size parameter $L = 12$ and truncation radius $R = 7.5$. The collision operators are applied to the Wigner state in \eqref{eq:validationW} and compared to a reference calculation with $J, M, N = 72$, respectively. The relative error is calculated using the $L^1$-norm of the representation as $N \times N$ matrices in Fourier grid space. In (a) we have additionally fixed $N = M = 32$, in (b) $J = 32$ and in (c) $J = 32$, $M = 72$.}
\end{figure}

First we change $J$ while keeping the other parameters fixed, and compare the result to a reference calculation with large $J = 72$. The relative error is calculated as $\sum_{j=0}^3 \lVert\mathcal{C}_{\mathrm{d}}[W]_j - \mathcal{C}_{\mathrm{d},\mathrm{ref}}[W]_j\rVert / \lVert \mathcal{C}_{\mathrm{d},\mathrm{ref}}[W]_j\rVert$ (and correspondingly for $\mathcal{C}_{\mathrm{c}}$), where $j$ denotes the Pauli matrix component as in \eqref{eq:spin_repr} and $\lVert \cdot \rVert$ is the $L^1$-norm of the representation as $N \times N$ matrices in Fourier grid space. One observes in Figure~\ref{fig:conv-J} that the relative error decreases exponentially fast as $J$ increases, confirming the previous discussion of exponential convergence with respect to $J$. Figure~\ref{fig:conv-J} also illustrates that $J = 32$ already achieves relative errors smaller than $10^{-8}$ (note that $J=32$ amounts to $64$ grid points in $\theta$ due to symmetry). We will fix this choice in the sequel.

Next, we study the accuracy of the approximation \eqref{eq:approxH-JM} for different choices of $M$. As discussed before, the integrand becomes more oscillatory as $N$ increases, and thus we expect that the number of quadrature nodes $M$ depends linearly on $N$. The relative error of the dissipative $\mathcal{C}_{\mathrm{d}}[W]$ collision term also seems to decay exponentially with $M$ before reaching the machine accuracy (see Figure~\ref{fig:conv-M}),  thanks to the accuracy of the Gauss-Legendre quadrature. In our subsequent computations we fix $M = 32$, which is adequate to achieve $10^{-5}$ relative error for $N = 32$.

Finally, the exponential convergence with respect to $N$ in Figure~\ref{fig:conv-N} verifies that our method achieves spectral accuracy in dealing with the collision operators. A large $M = 72$ is used in Figure~\ref{fig:conv-N} to ensure the accuracy of the approximation \eqref{eq:approxH-JM} for each $N$.

\subsection{Spatially homogeneous equation}

\begin{figure}[!ht]
\centering
\subfloat[spin density matrix]{\includegraphics[width=0.3\textwidth]{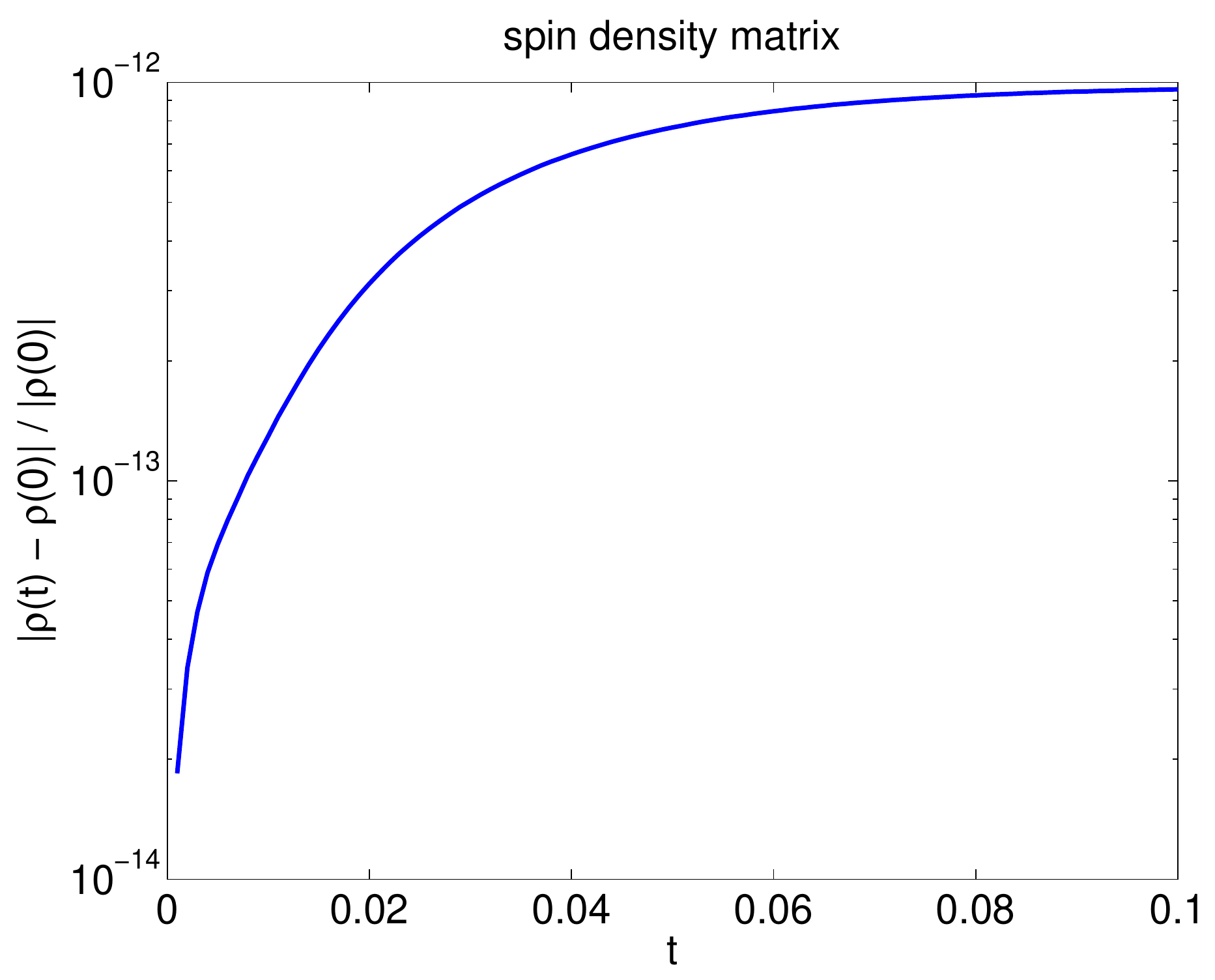}}
\hspace{0.04\textwidth}
\subfloat[momentum]{\includegraphics[width=0.3\textwidth]{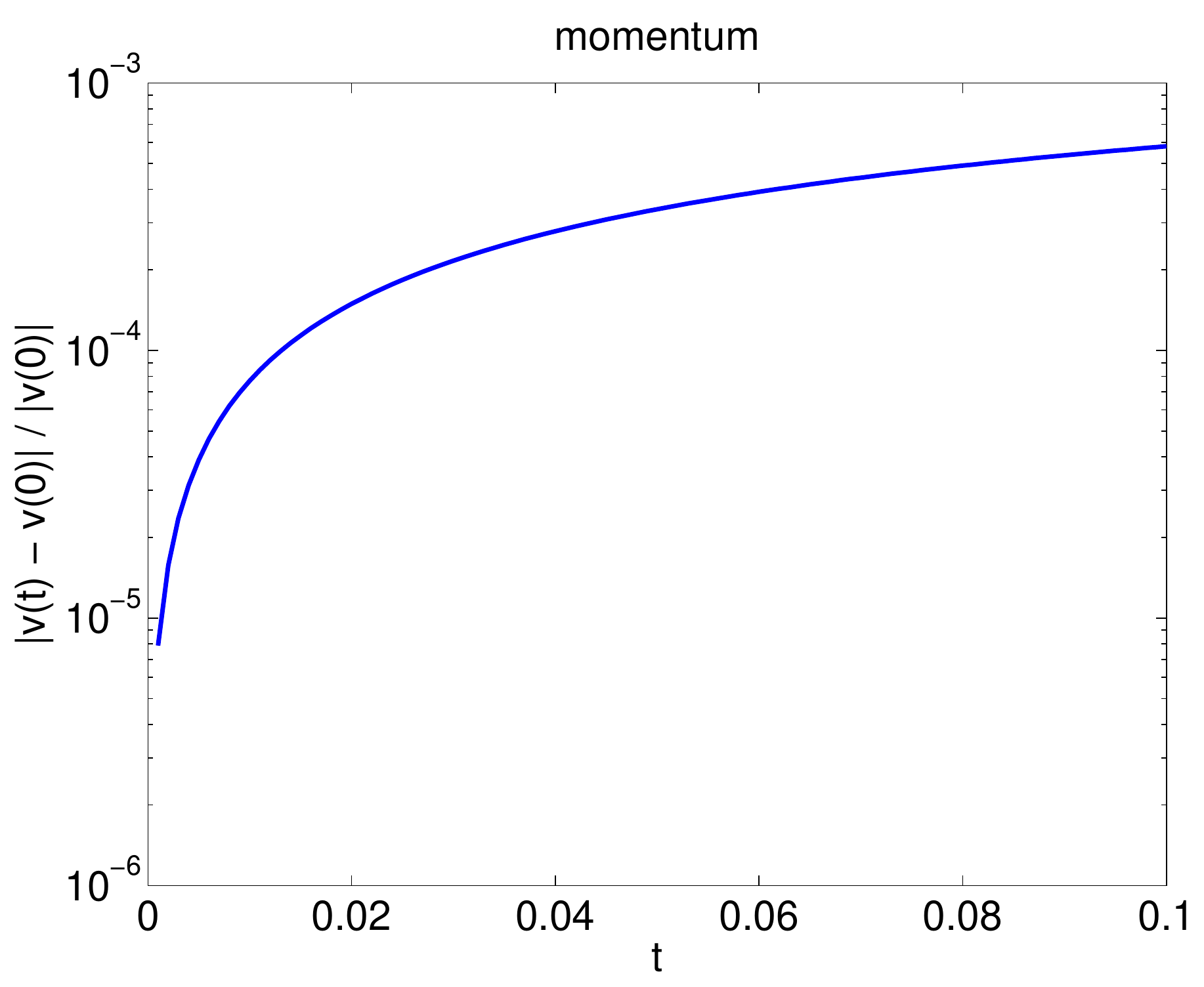}}
\hspace{0.04\textwidth}
\subfloat[energy]{\includegraphics[width=0.3\textwidth]{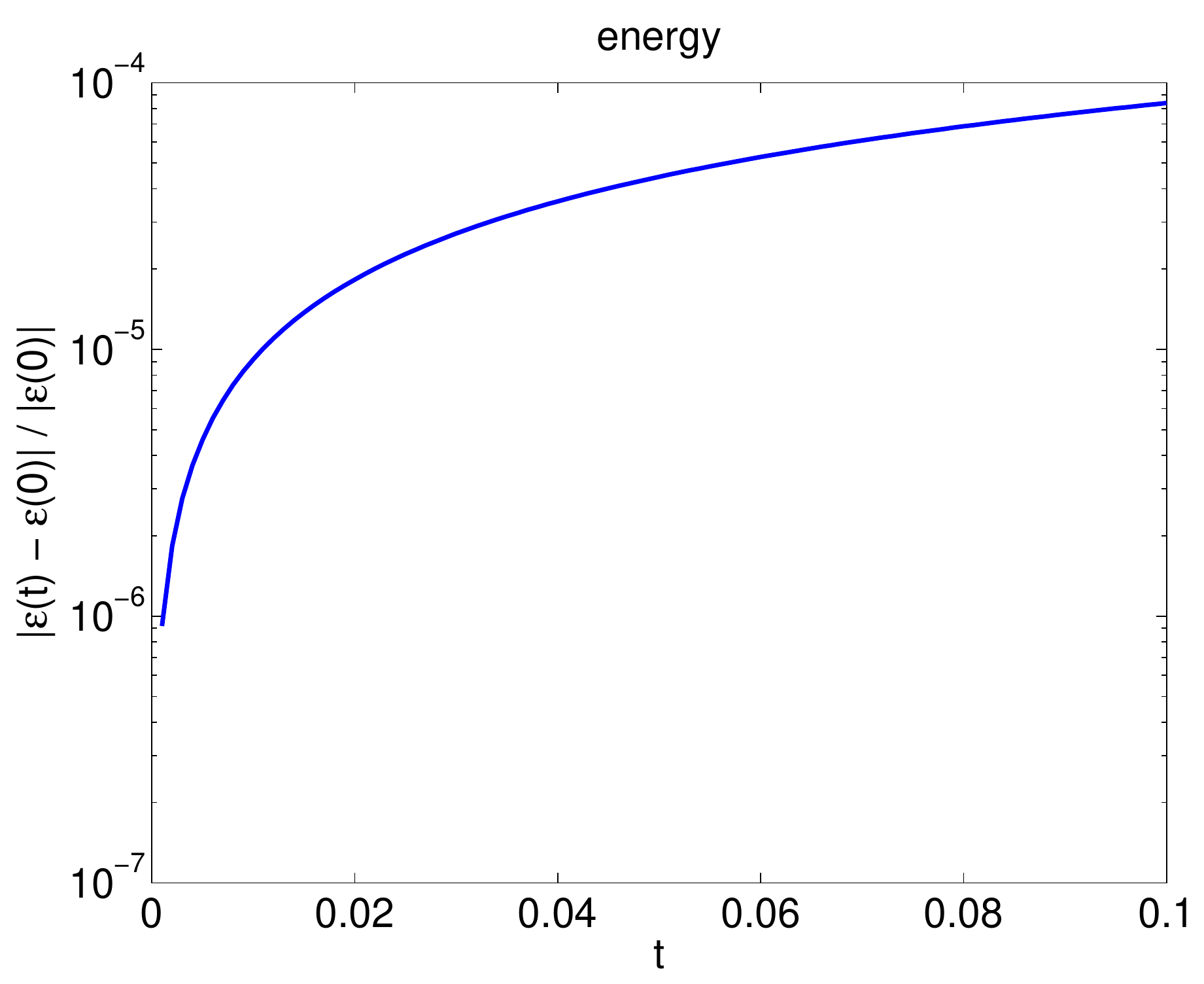}}
\caption{Conservation of the spin density \eqref{eq:density}, momentum \eqref{eq:velocity} and energy \eqref{eq:internal_energy} by the numerical scheme for the spatially homogeneous setting with parameters $N = 32$, $L = 12$, $J = 32$, $M = 32$, $R = 7.5$, initial state \eqref{eq:W0homogeneous} and time step $\Delta t = 0.001$. The y-axis shows the relative deviation from the initial value on a logarithmic scale.}
\label{fig:hom_conservation}
\end{figure}

We now study the time evolution under the spatially \emph{homogeneous} equation. Here we choose the initial condition to be a Fermi-Dirac state perturbed by $v$-dependent rotations:
\begin{equation}\label{eq:W0homogeneous}
W(0,v) = \mathrm{e}^{-i X(v)} \cdot U_0 \cdot \mathrm{diag}\big( \mathrm{e}^{\left(\frac{1}{2} \abs{v - v_0}^2 - \mu_s\right)/(k_{\mathrm{B}} T)} + 1 \big)^{-1}_s \cdot U_0^* \cdot \mathrm{e}^{i X(v)}
\end{equation}
with
\begin{equation*}
U_0 = \begin{pmatrix} \cos(\pi/5) & -i \sin(\pi/5) \\ \sin(\pi/5) & i \cos(\pi/5) \end{pmatrix}, \quad 
X(v) = \sin\big(v_y^2\big)\,\mathbbm{1} + (v_x-1)\sigma_1 + 2 \big(v_x^2 + \abs{v_y} + 1\big)^{-1} \sigma_2 + \cos\big(v_x + \tfrac{1}{2}v_y\big)\,\sigma_3
\end{equation*}
and the parameters $k_{\mathrm{B}} T = 5/4$, $\mu_{\uparrow} = 1$, $\mu_{\downarrow} = 3/2$ and $v_0 = (0.4, -0.1)$. In Figure~\ref{fig:hom_conservation}, the conserved quantities spin density \eqref{eq:density}, momentum \eqref{eq:velocity} and energy \eqref{eq:internal_energy} are plotted as a function of time. We use the matrix $L^2$-norm for the relative error of the $2 \times 2$ spin density $\rho(t)$. Excellent conservation is observed numerically (note that the scale of the y-axes are $10^{-12}$, $10^{-4}$ and $10^{-5}$ respectively). We have used the time step $\Delta t = 0.001$ for the simulation.
\begin{figure}[!ht]
\centering
\includegraphics[width=0.35\textwidth]{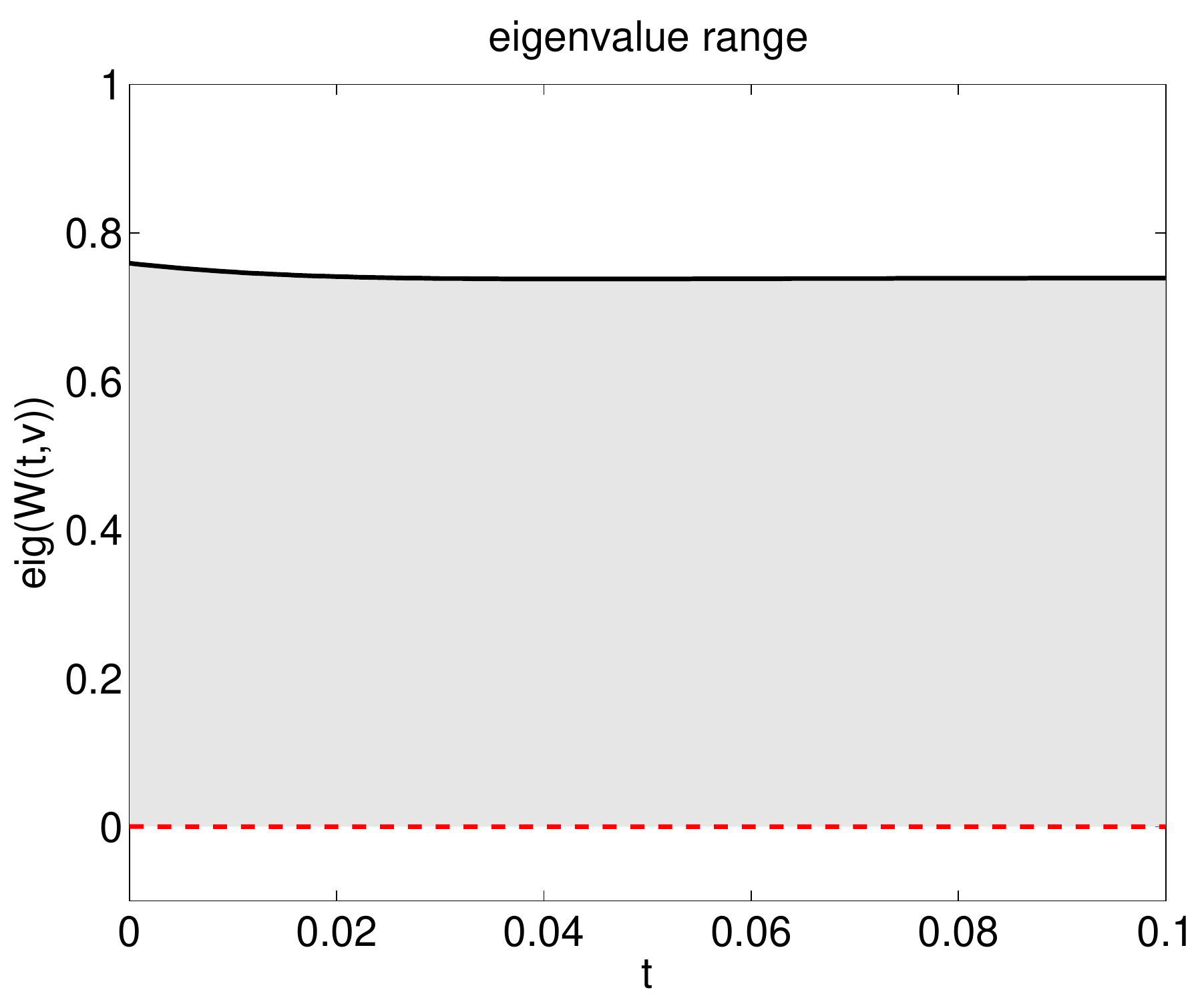}
\caption{Time-dependent range of the eigenvalues of the Wigner states, remaining between $0$ and $1$, as required (same simulation as in Figure~\ref{fig:hom_conservation}).}
\label{fig:eigenvalue_range}
\end{figure}
As another consistency check, the eigenvalues of the Wigner spin-density matrices must stay between $0$ and $1$. This condition is satisfied by our numerical scheme: Figure~\ref{fig:eigenvalue_range} visualizes the largest and smallest (with respect to $v$) of all eigenvalues.
\begin{figure}[!ht]
\centering
\subfloat[entropy]{\label{fig:entropy}\includegraphics[width=0.3\textwidth]{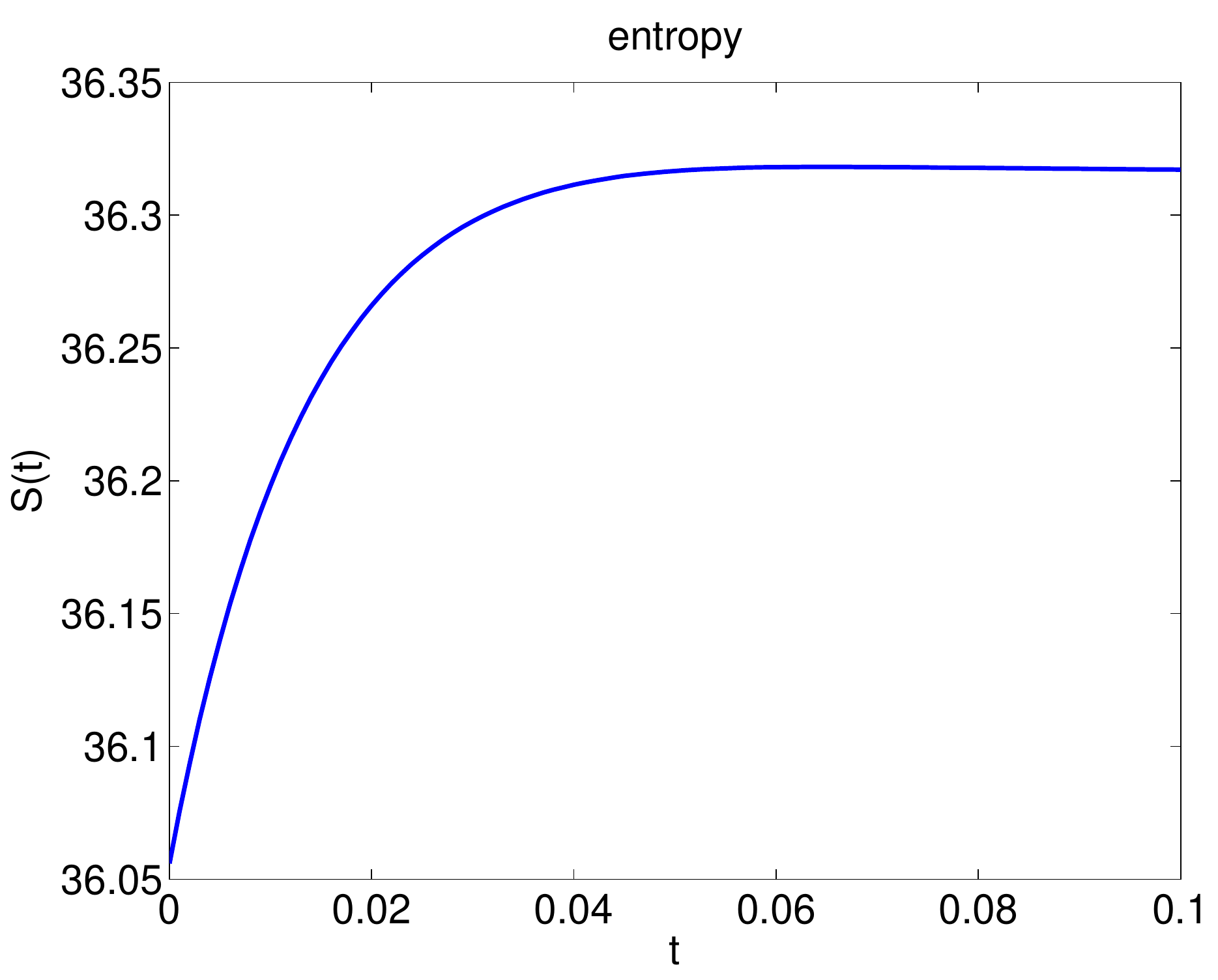}}
\hspace{0.04\textwidth}
\subfloat[quantum relative entropy]{
\label{fig:relentropy}
\includegraphics[width=0.3\textwidth]{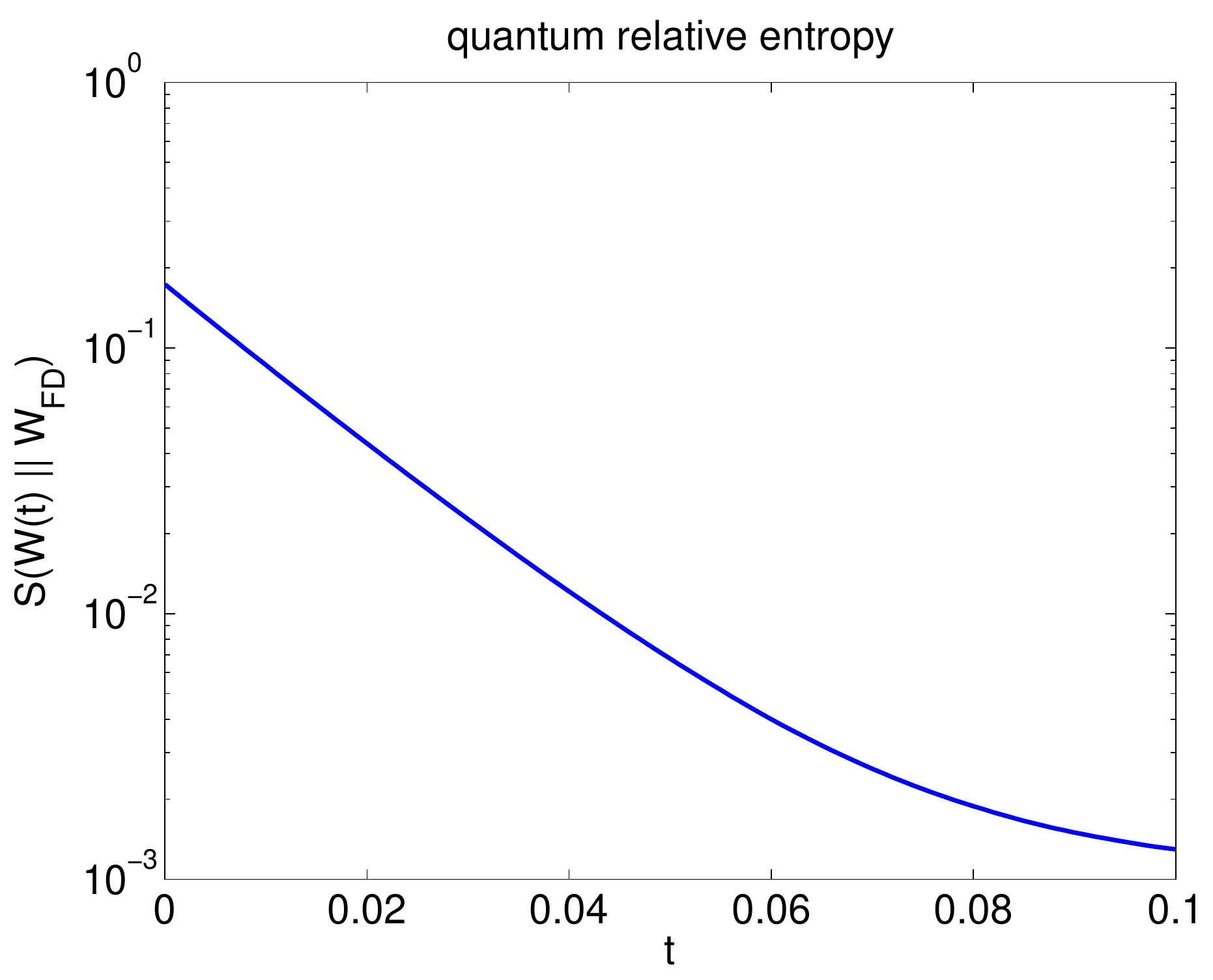}}
\hspace{0.04\textwidth}
\subfloat[convergence to $W_{\FD}$]{
\label{fig:convergenceFD}
\includegraphics[width=0.3\textwidth]{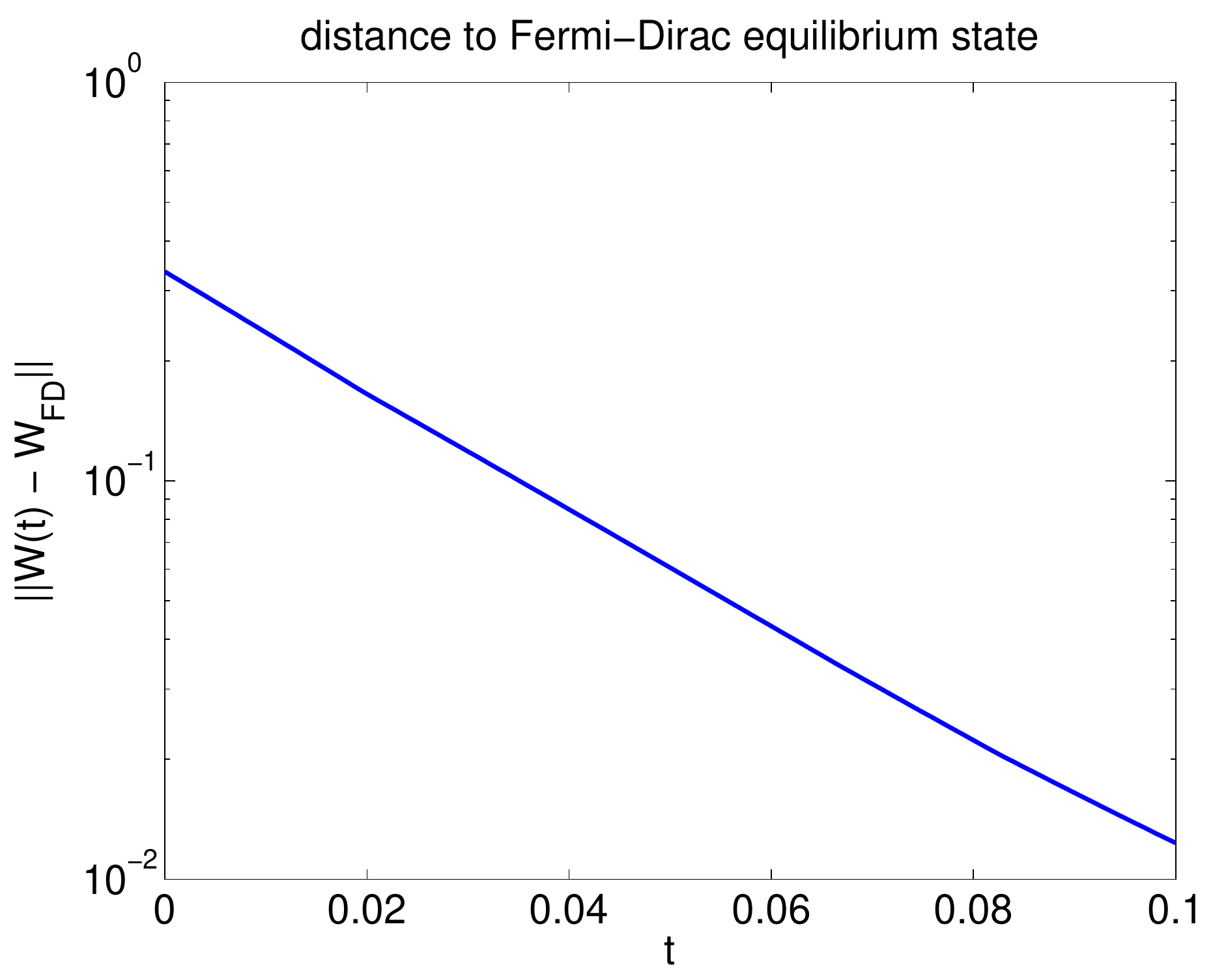}}
\caption{Time-dependent entropy (a) and quantum relative entropy (b) as specified in Eq.~\eqref{eq:relentropy_hom}, as well as the $L^1$-norm distance (c) to the asymptotic Fermi-Dirac equilibrium state determined by the conservation laws (spatially homogeneous case, same simulation as in Figure~\ref{fig:hom_conservation}).}
\end{figure}
In accordance with the H-theorem, the entropy is monotonically increasing (see Figure~\ref{fig:entropy}). Physically, as $t \to \infty$ the Wigner state should converge to a thermal equilibrium Fermi-Dirac distribution $W_{\FD}$ \eqref{eq:W_FermiDirac} with moments \eqref{eq:momentsFermiDirac} matching the conserved moments of the Wigner state (see Figure~\ref{fig:hom_conservation}). The average velocity $u$ leads to a shift $W_{\FD}(v) \to W_{\FD}(v - u)$ in \eqref{eq:W_FermiDirac}, and the eigenbasis of $W_{\FD}$ is equal to the eigenbasis of the spin density matrix $\rho$. We fit the temperature $T$ and chemical potentials $\mu_{\uparrow}$, $\mu_{\downarrow}$ in \eqref{eq:momentsFermiDirac} numerically to match the eigenvalues of $\rho$ and the energy. The expected convergence to $W_{\FD}$ is verified in Figure~\ref{fig:relentropy}, showing the quantum relative entropy between the Wigner state and $W_{\FD}$. The quantum relative entropy is given by (recall that $\wt{W} = \mathbbm{1} -  W$)
\begin{equation}
\label{eq:relentropy_hom}
  S(W \parallel W_{\FD}) = \int \tr  \left[ W(v) \bigl(\log W(v) - \log W_{\FD}(v)\bigr) + \wt{W}(v) \bigl(\log \wt{W}(v) - \log \wt{W}_{\FD}(v) \bigr) \right]\ud v. 
\end{equation}
Using the H-theorem for the entropy and the conservation property of the fluid dynamic moments, the relative entropy is monotonically decaying to zero as $t \to \infty$. Finally, we demonstrate exponential convergence to $W_{\FD}$ in $L^1$-norm in Figure~\ref{fig:convergenceFD}.

\subsection{Spatially inhomogeneous equation with periodic boundary conditions}

\begin{figure}[!ht]
\centering
\subfloat[spin density matrix]{\includegraphics[width=0.3\textwidth]{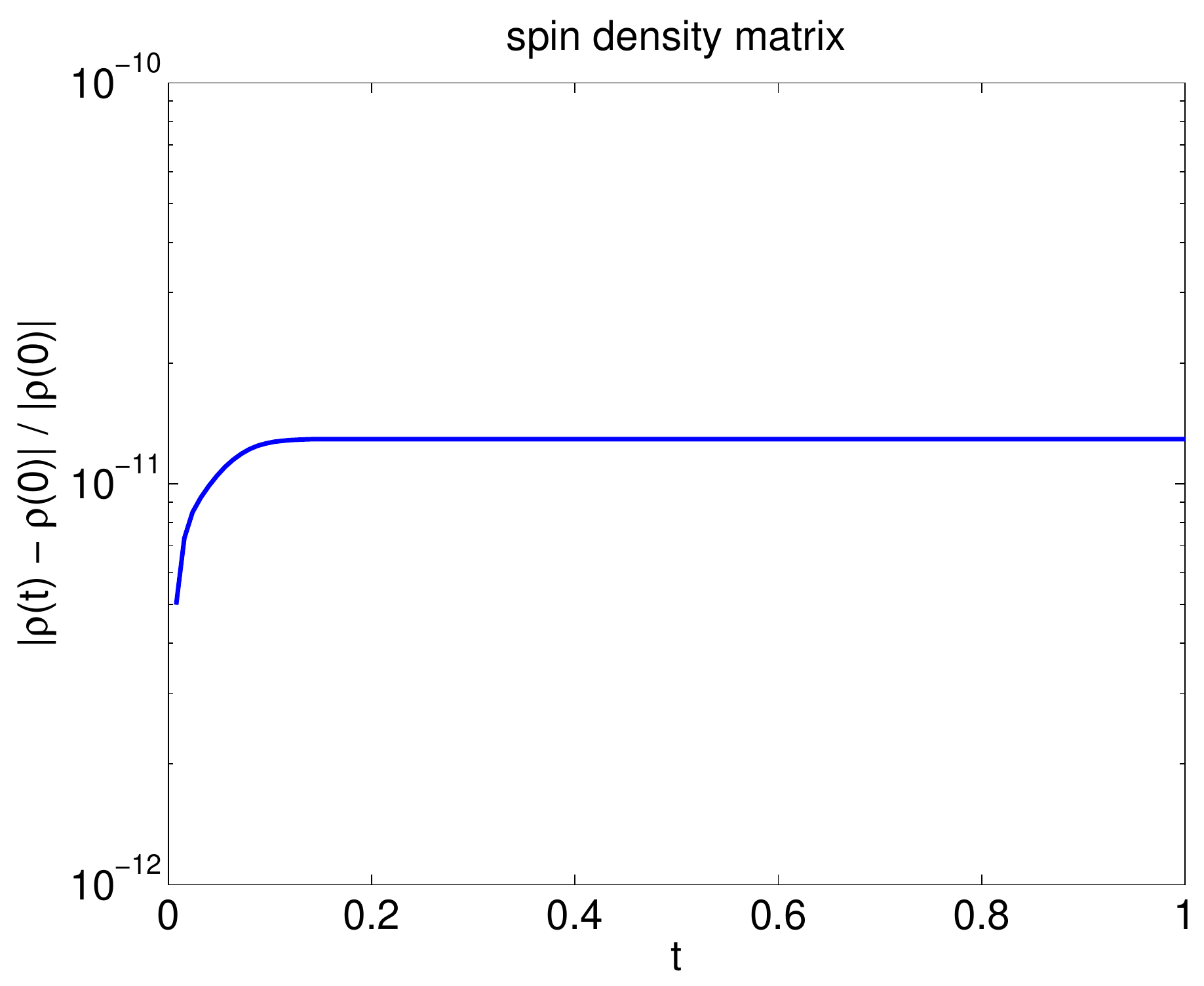}} \hspace{0.04\textwidth}
\subfloat[momentum]{\includegraphics[width=0.3\textwidth]{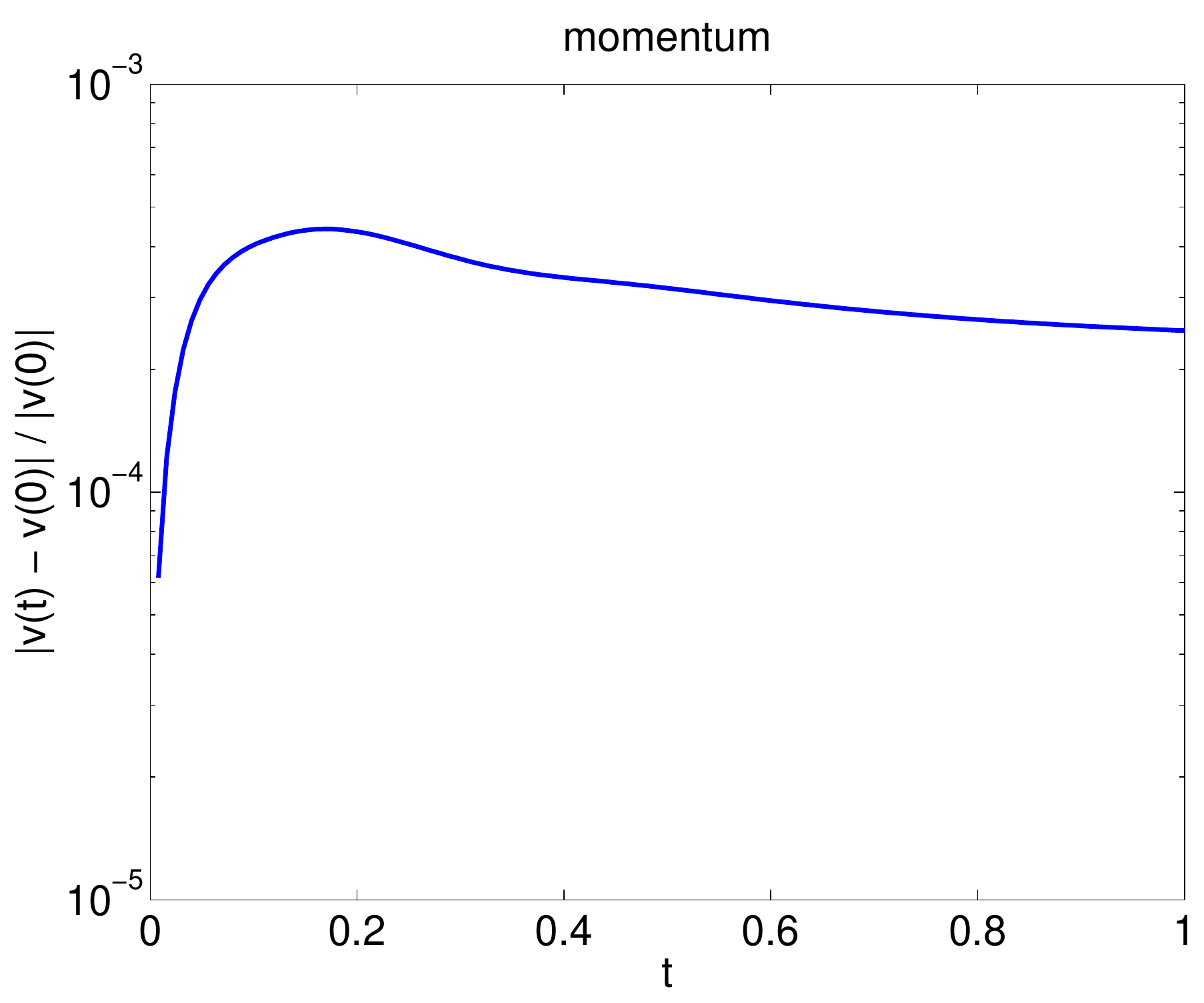}}
\hspace{0.04\textwidth}
\subfloat[energy]{\includegraphics[width=0.3\textwidth]{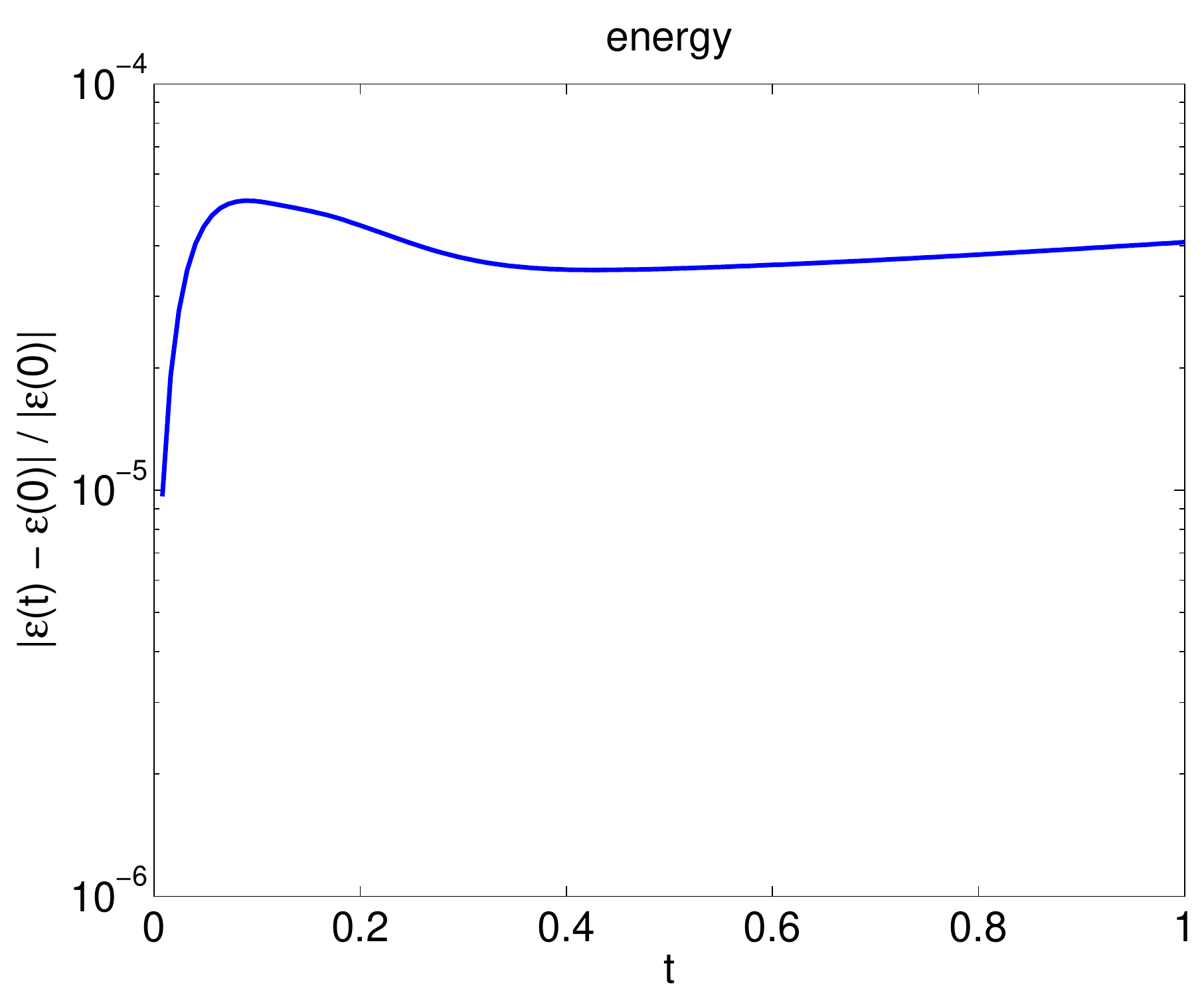}}
\caption{Conservation of the (spatially averaged) spin density \eqref{eq:density}, momentum \eqref{eq:velocity} and energy \eqref{eq:internal_energy} by the numerical scheme for the spatially inhomogeneous equation with periodic boundary conditions and $x \in [0,1)$. The parameters in this simulation are $N = 32$, $L = 12$, $J = 32$, $M = 32$, $R = 7.5$, the mesh width $\Delta x = 0.1$ and the time step $\Delta t = 0.008$.}
\label{fig:inhom_periodic_conservation}
\end{figure}

For \emph{periodic} boundary conditions, the spin density, momentum, and energy are still conserved globally, i.e., after taking the integral over the spatial dimension. Figure~\ref{fig:inhom_periodic_conservation} shows the conservation in the numerical scheme, analogous to the spatially homogeneous case. The initial Wigner state is similar to \eqref{eq:W0homogeneous} but with $T$, $\mu_s$ and $v_0$ depending on the spatial location $x$.

\begin{figure}[!ht]
\centering
\subfloat[entropy]{\label{fig:inhom_periodic_entropy}\includegraphics[width=0.35\textwidth]{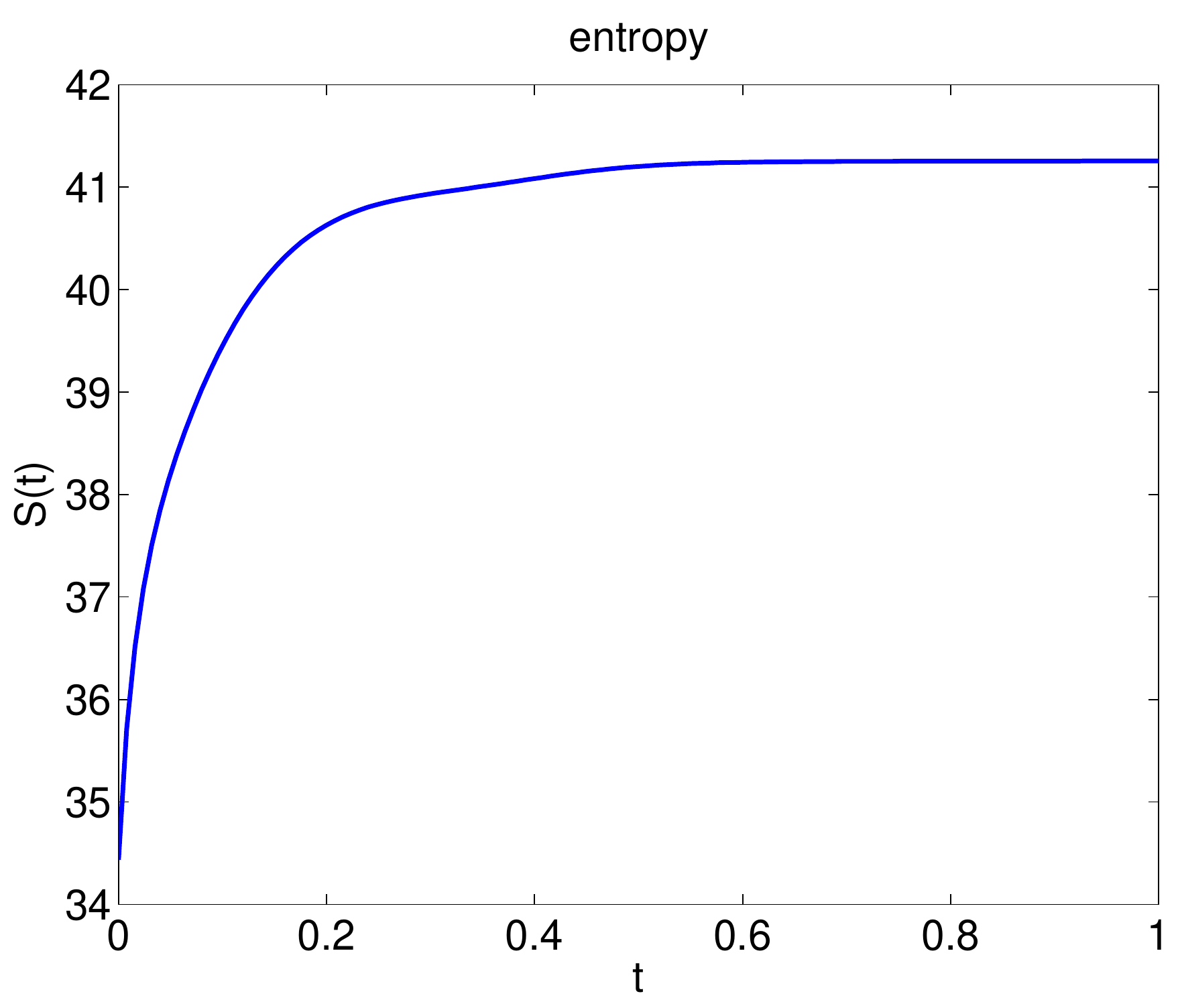}}
\hspace{0.1\textwidth}
\subfloat[quantum relative entropy]{\label{fig:inhom_periodic_relentropy}\includegraphics[width=0.35\textwidth]{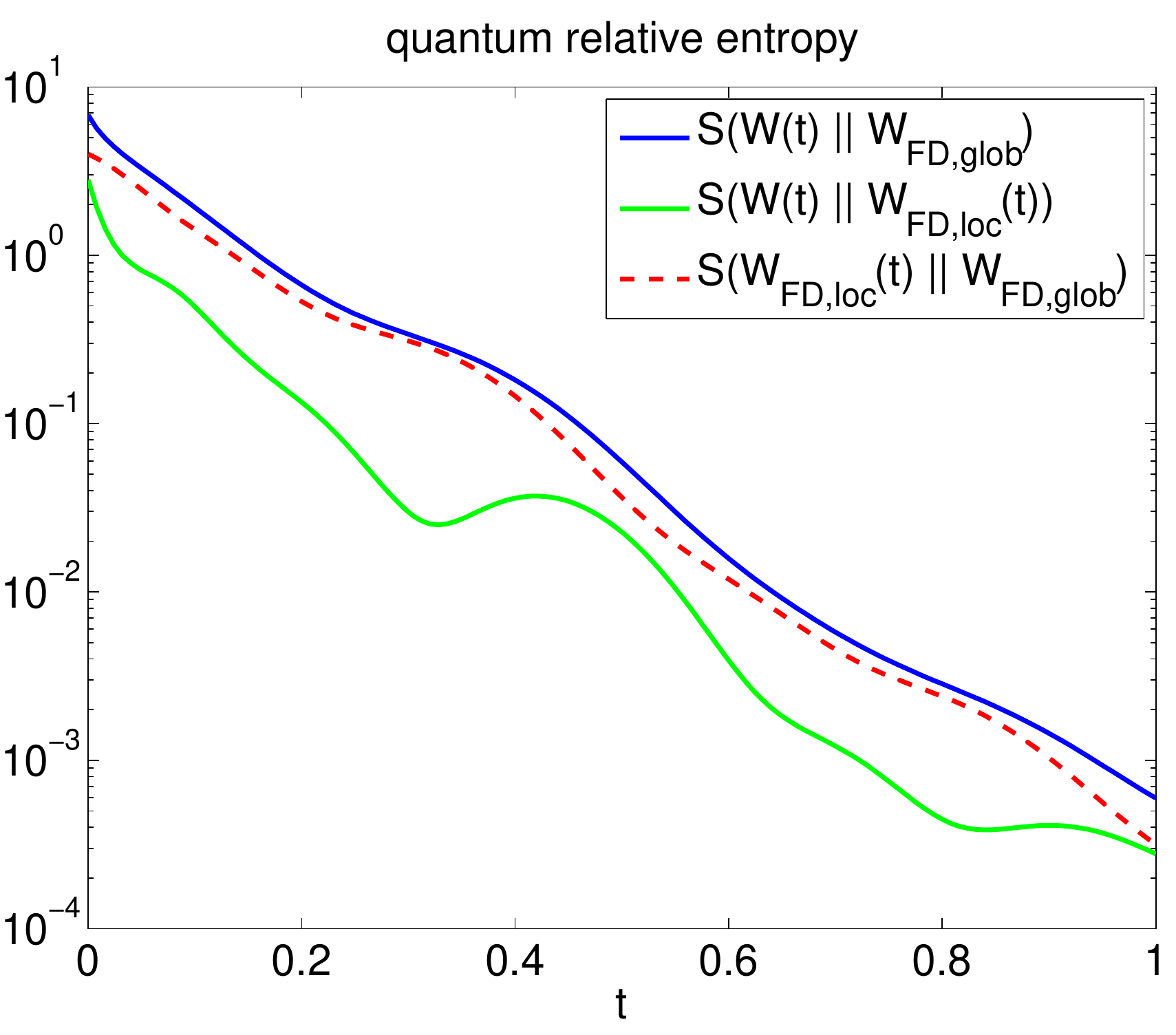}}
\caption{Time-dependent entropy (a) and quantum relative entropy (b) for the spatially inhomogeneous equation with periodic boundary conditions and $x \in [0,1)$. In (b), the blue curve shows the entropy relative to the global, uniform Fermi-Dirac state (Eq.~\eqref{eq:relentropy_inhom}), whereas the green curve displays the entropy relative to locally fitted (at each $x$ and $t$) Fermi-Dirac states (Eq.~\eqref{eq:relentropy_local}).} 
\end{figure}

The quantum relative entropy now involves integration over the spatial domain:
\begin{equation}
\label{eq:relentropy_inhom}
S\big(W(t,\cdot,\cdot) \parallel W_{\FD}\big) = \int S\big(W(t,x,\cdot) \parallel W_{\FD}\big) \ud x. 
\end{equation}
Note that the asymptotic Fermi-Dirac equilibrium state is independent of $t$ and $x$. For comparison, we define a quantum relative entropy with respect to locally fitted (at each $x$ and $t$) Fermi-Dirac states:
\begin{equation}
\label{eq:relentropy_local}
S_{\mathrm{loc}}\big(W(t,\cdot,\cdot) \parallel W_{\FD,\mathrm{loc}}(t,\cdot,\cdot)\big) = \int S\big(W(t,x,\cdot) \parallel W_{\FD,\mathrm{loc}}(t,x,\cdot)\big) \ud x. 
\end{equation}
Here $W_{\FD,\mathrm{loc}}(t,x,\cdot)$ is the Fermi-Dirac state with the same spin density, momentum and energy as $W(t,x,\cdot)$. Figure~\ref{fig:inhom_periodic_entropy} visualizes the monotonically increasing entropy, and Figure~\ref{fig:inhom_periodic_relentropy} the quantum relative entropy, both global and local (for the same simulation as in Figure~\ref{fig:inhom_periodic_conservation}). It is straightforward to verify that 
\begin{equation}
  S\big(W(t) \parallel W_{\FD}\big) = S\big(W(t) \parallel W_{\FD, \mathrm{loc}}(t)\big) 
+ S\big(W_{\FD, \mathrm{loc}}(t) \parallel W_{\FD}\big).
\end{equation}
According to Figure~\ref{fig:inhom_periodic_relentropy}, the quantum relative entropy is dominated by $S\big(W_{\FD, \mathrm{loc}}(t) \parallel W_{\FD}\big)$. In other words, the system quickly relaxes to a local Fermi-Dirac state, before converging to the global equilibrium.

\smallskip

As next step, we investigate the effect of an external, $x$-dependent magnetic field, which enters the Boltzmann equation as the last term in Eq.~\eqref{eq:boltzmann}. Specifically, we choose
\begin{equation}\label{eq:Bext_rot}
\vec{B}(x) = (0, \cos(2\pi x), \sin(2\pi x))^T
\end{equation}
for the simulation.
\begin{figure}[!ht]
\centering
\subfloat[external magnetic field]{\includegraphics[width=0.3\textwidth]{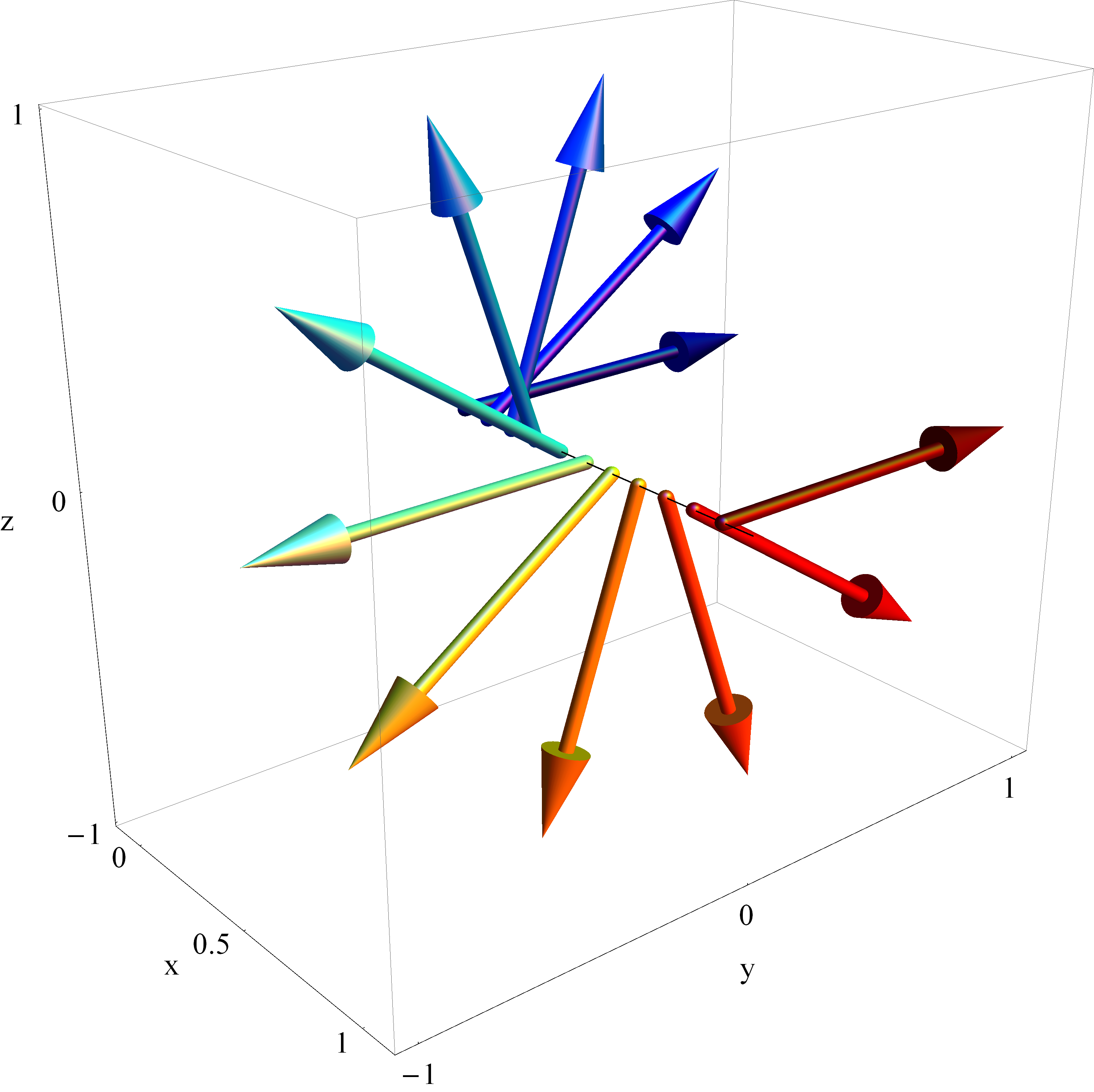}} \hspace{0.04\textwidth}
\subfloat[density $\rho(t,x)$ at $t = 1$]{\includegraphics[width=0.3\textwidth]{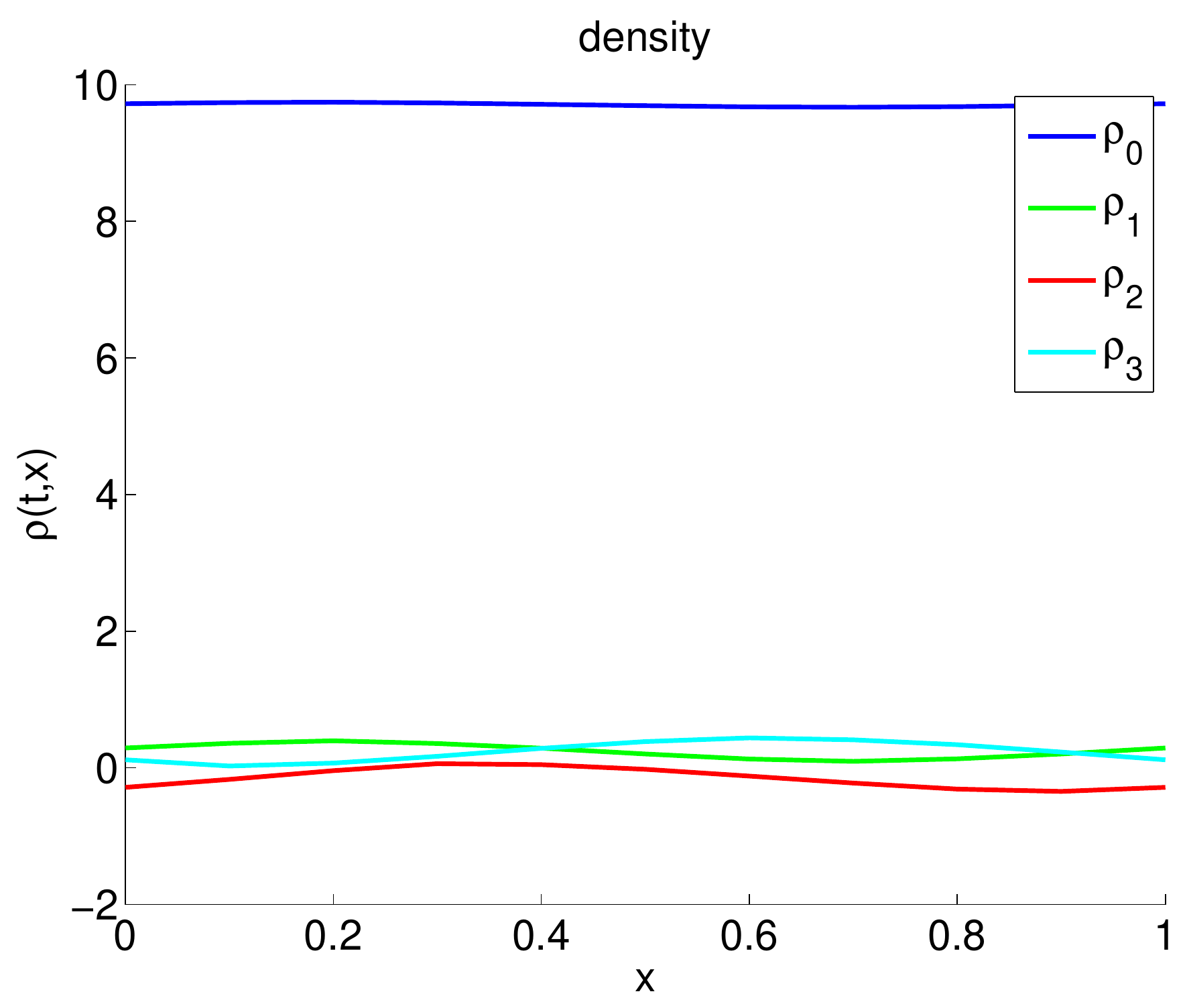}} \hspace{0.04\textwidth}
\subfloat[$\tfrac{1}{2}$ Bloch vector of density]{\includegraphics[width=0.3\textwidth]{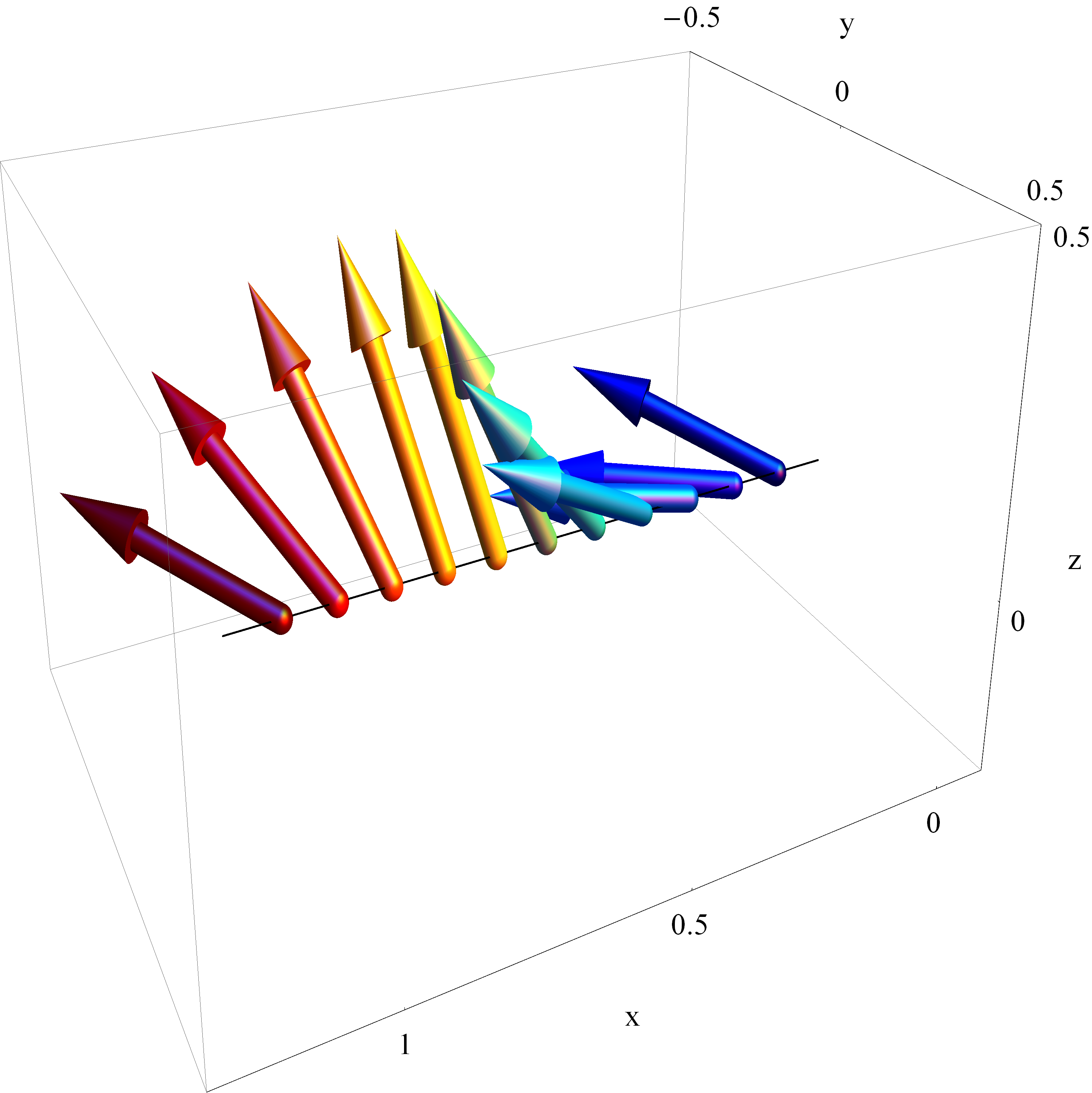}}
\caption{(a) Visualization of the external magnetic field in Eq.~\eqref{eq:Bext_rot}, with color encoding the position along the $x$ axis. (b) Components of the corresponding spin density matrix $\rho(t,x)$ in the Pauli representation \eqref{eq:spin_repr} for a simulation with this magnetic field, periodic boundary conditions and finite volume size $\Delta x = 0.1$. (c) Visualization of the corresponding Bloch vector part of the density (for clarity, the axis is rotated compared to (a)).}
\label{fig:density_Bext_rot}
\end{figure}
Figure~\ref{fig:density_Bext_rot} visualizes the external magnetic field and shows the components of the spin density matrix at $t = 1$, with the Bloch vector part (scaled by $\frac{1}{2}$) defined as $\vec{\rho} = (\rho_1, \rho_2, \rho_3)$. Compared to the above simulation without magnetic field (density not shown), the Bloch vector components of $\rho(t,x)$ now change with $x$. Since the magnetic field acts as a unitary rotation of the Wigner state in the time evolution, the trace $\tr[W(t,x,v)]$ remains unaffected by the magnetic field and the momentum and energy conservation laws still hold.

\subsection{Spatially inhomogeneous equation with Dirichlet and Maxwell boundary conditions}

First, we investigate a simulation with Dirichlet boundary conditions. The fixed states at the left and right boundary are Fermi-Dirac states \eqref{eq:W_FermiDirac} with different temperatures and eigenbasis, as summarized in Table~\ref{tab:Dirichlet}.
\begin{table}[!ht]
\begin{tabular}{r|cc}
& left & right \\
\hline
$1/(k_{\mathrm{B}} T)$ & 0.8 & 1.2 \\
$\mu_{\uparrow}$ & 1.5 & 1.5 \\
$\mu_{\downarrow}$ & -1.5 & -1.5 \\
$U$ & $\mathbbm{1}$ & $\frac{1}{\sqrt{2}} \left(\begin{smallmatrix}1 & -1 \\ 1 & 1\end{smallmatrix}\right)$ \\
\hline
\end{tabular}
\smallskip
\caption{Temperature $T$, chemical potentials $\mu_{\uparrow}$, $\mu_{\downarrow}$ and eigenbasis $U$ of the incoming left and right Fermi-Dirac boundary states for a simulation with Dirichlet boundary conditions.}
\label{tab:Dirichlet}
\end{table}
Here, $U$ contains the spin-eigenbasis of the Fermi-Dirac state as column vectors.
\begin{figure}[!ht]
\centering
\subfloat[stationary density]{\includegraphics[width=0.3\textwidth]{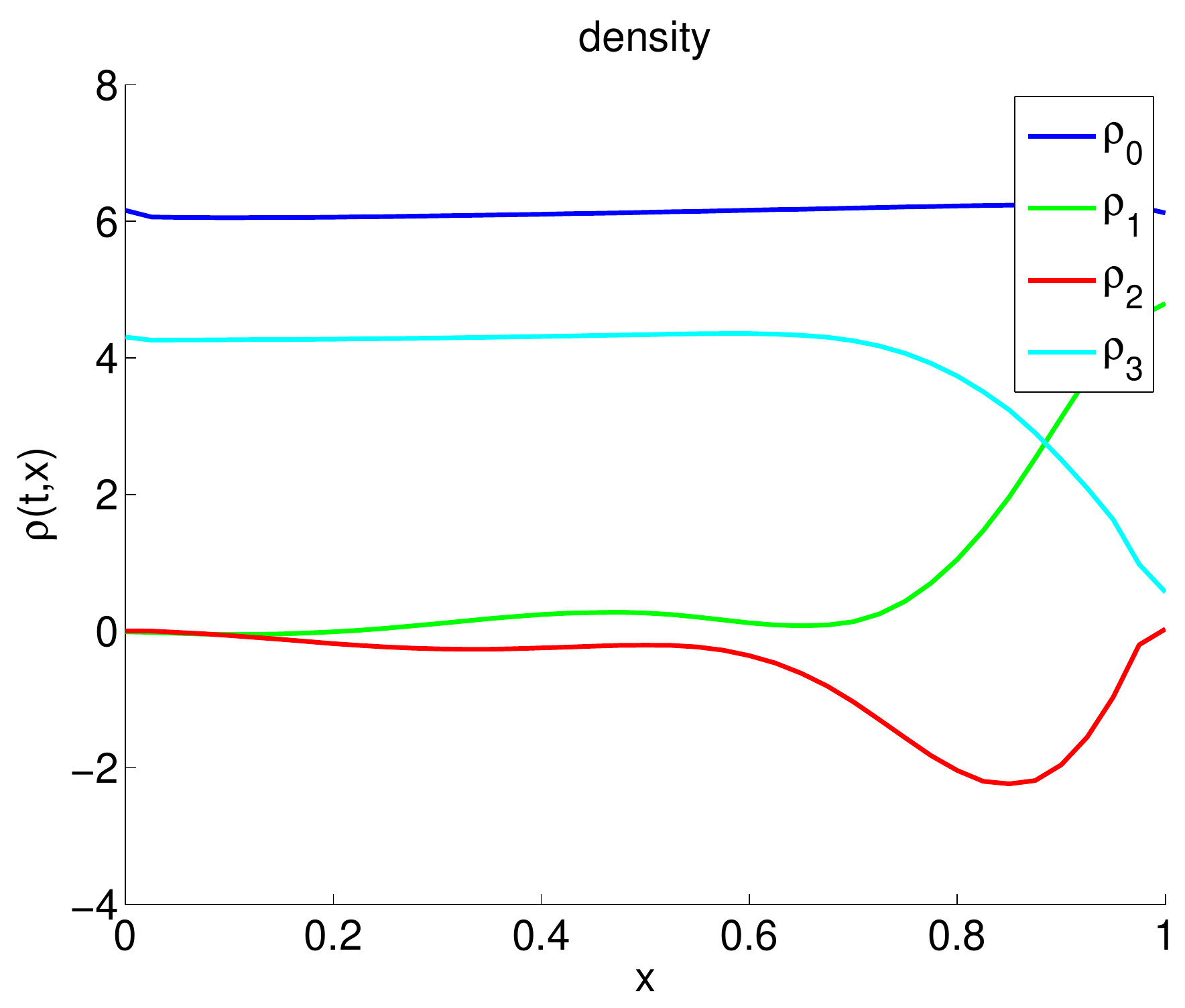}} \hspace{0.04\textwidth}
\subfloat[$k_{\mathrm{B}} T$]{\includegraphics[width=0.3\textwidth]{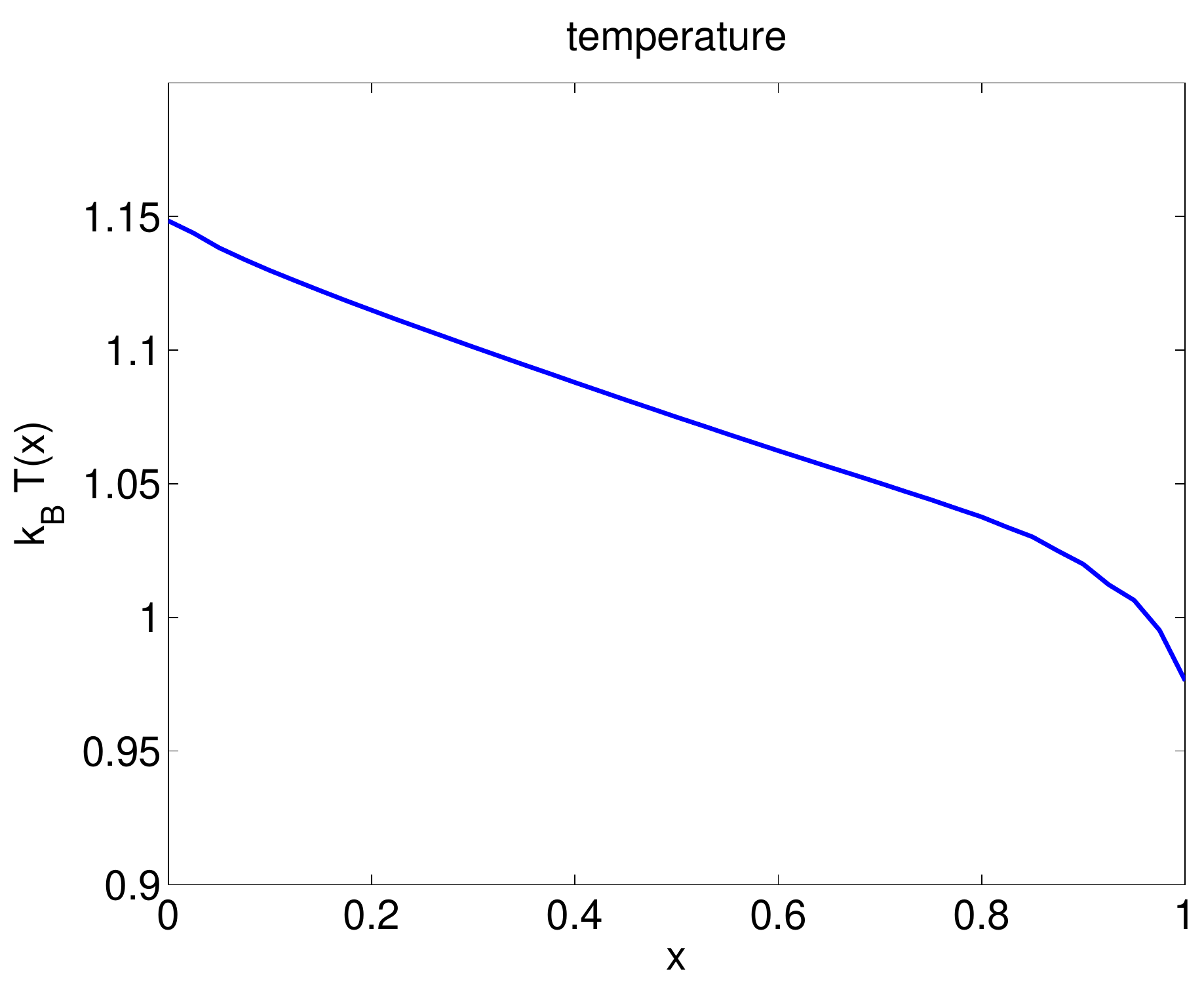}} \hspace{0.04\textwidth}
\subfloat[local entropy]{\includegraphics[width=0.3\textwidth]{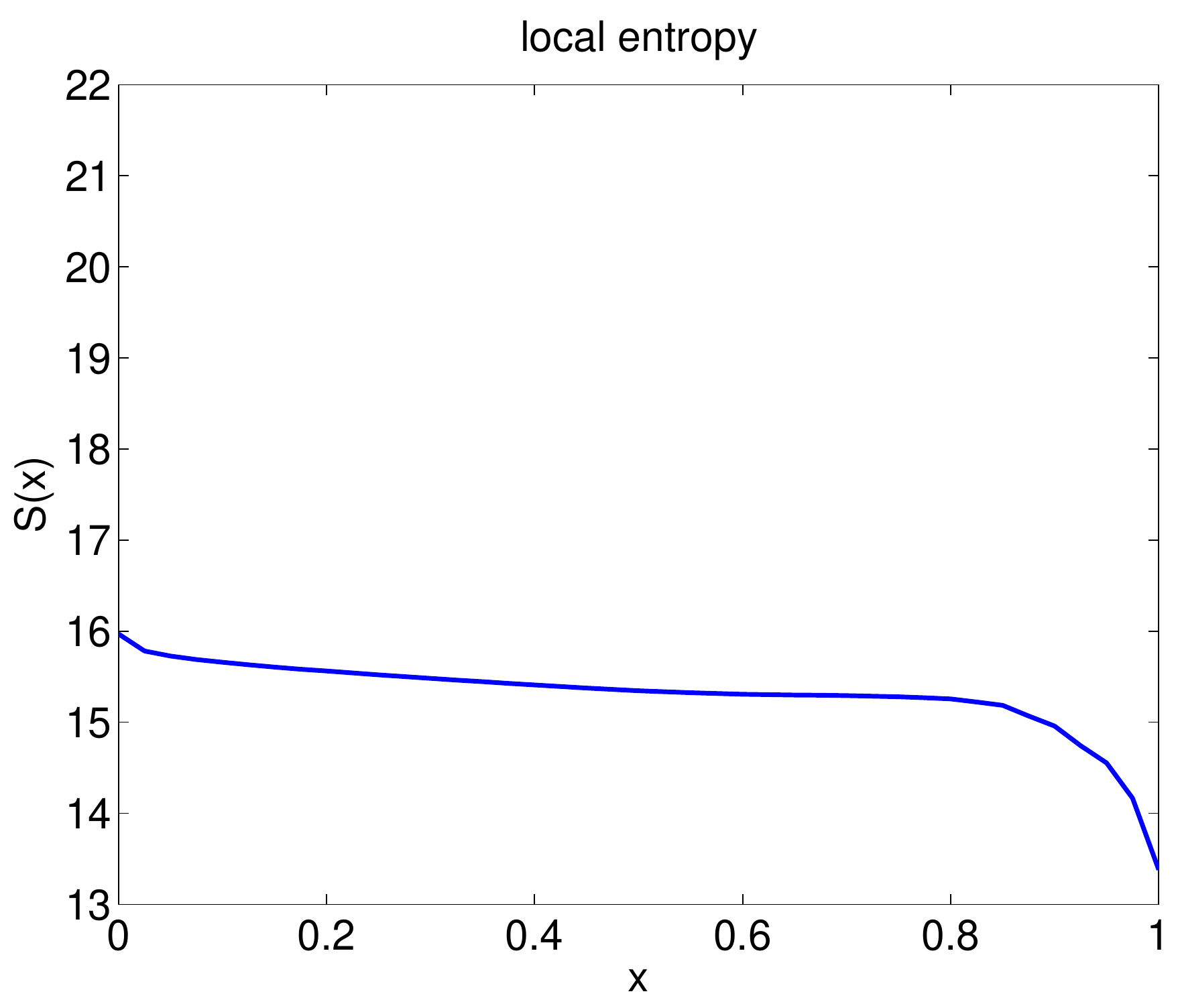}}
\caption{Components of the stationary spin density matrix (A) in the Pauli representation \eqref{eq:spin_repr}, local temperature (B) and local entropy (C) for a simulation with Dirichlet boundary conditions, with finite volume size $\Delta x = 0.025$.}
\label{fig:Dirichlet}
\end{figure}
In Figure~\ref{fig:Dirichlet} we visualize the stationary density (after running the simulation until reaching stationarity) as well as the local temperature and entropy. The temperature is estimated by constructing a local Fermi-Dirac state with the same moments as the actual local Wigner state. Note that the stationary temperature at the boundary is not exactly equal to the values in Table~\ref{tab:Dirichlet} since the Dirichlet boundary condition fixes only the incoming parts of the Wigner states at the left and right boundary.

\medskip

Finally, we explore the effects of Maxwell boundary states with (i) different temperature and chemical potentials but common spin eigenbasis, and (ii) same temperature and chemical potentials but spin eigenvectors pointing to different directions, as summarized in Table~\ref{tab:Maxwell}.
\begin{table}[!ht]
\begin{tabular}{r|cc|cc|cc|cc}
& \multicolumn{2}{|c|}{$1/(k_{\mathrm{B}} T)$} & \multicolumn{2}{|c}{$\mu_{\uparrow}$} & \multicolumn{2}{|c}{$\mu_{\downarrow}$} & \multicolumn{2}{|c}{$U$} \\
& left & right & left & right & left & right & left & right \\
\hline
Maxwell (i)  & 0.6 & 1.1 & -0.4 & 1.3 & 1.8 & -0.9 & $\mathbbm{1}$ & $\mathbbm{1}$ \\
Maxwell (ii) & 1 & 1 & 1.5 & 1.5 & -1.5 & -1.5 & $\mathbbm{1}$ & $\frac{1}{\sqrt{2}} \left(\begin{smallmatrix}1 & -1 \\ 1 & 1\end{smallmatrix}\right)$ \\
\hline
\end{tabular}
\smallskip
\caption{Temperature $T$, chemical potentials $\mu_{\uparrow}$, $\mu_{\downarrow}$ and eigenbasis $U$ of the left and right diffusive reflection Fermi-Dirac states for two simulations with Maxwell boundary conditions.}
\label{tab:Maxwell}
\end{table}
\begin{figure}[!ht]
\centering
\subfloat[stationary density]{
\label{fig:Maxwell1_density}
\includegraphics[width=0.3\textwidth]{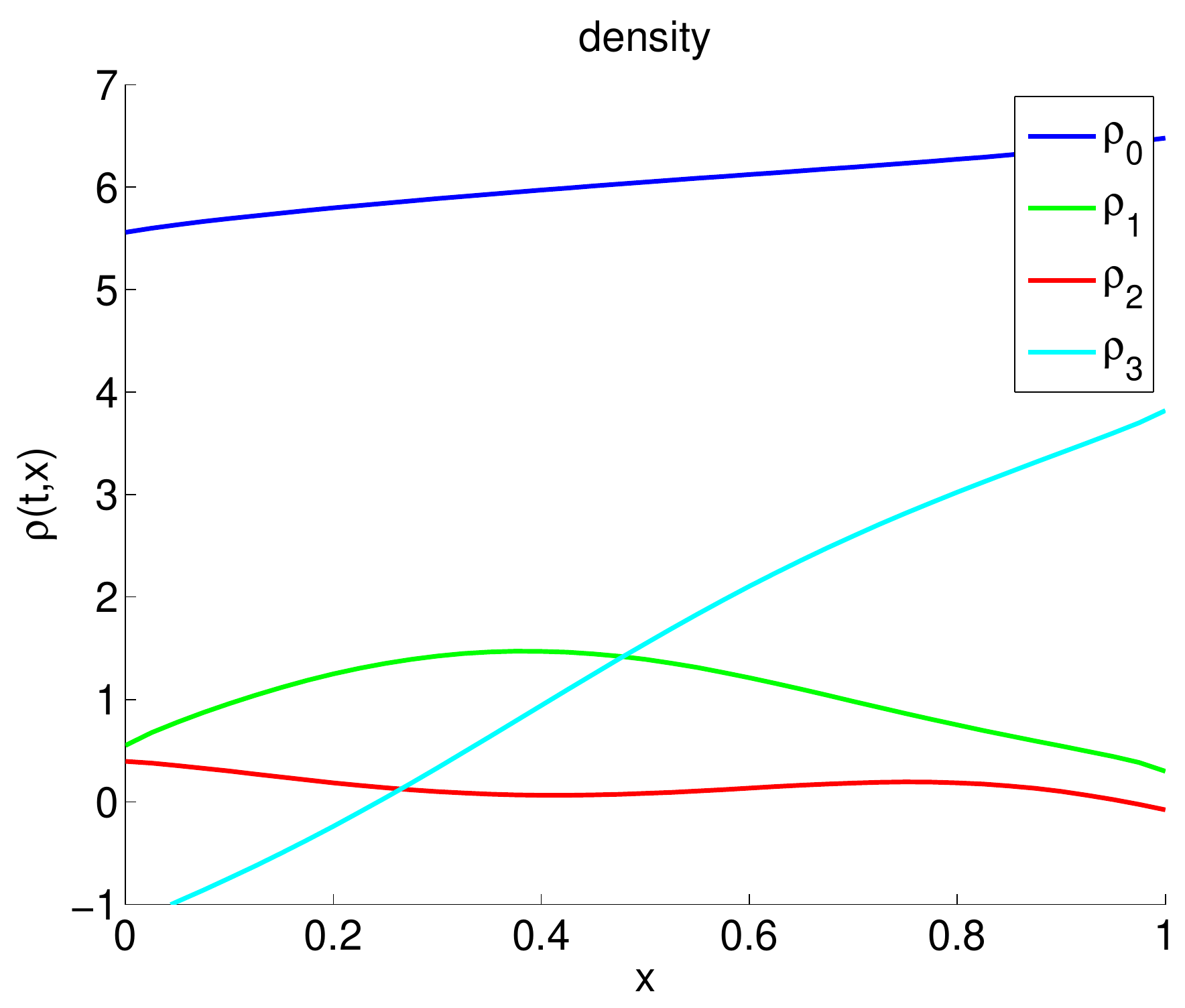}} \hspace{0.04\textwidth}
\subfloat[$k_{\mathrm{B}} T$]{
\label{fig:Maxwell1_temperature}
\includegraphics[width=0.3\textwidth]{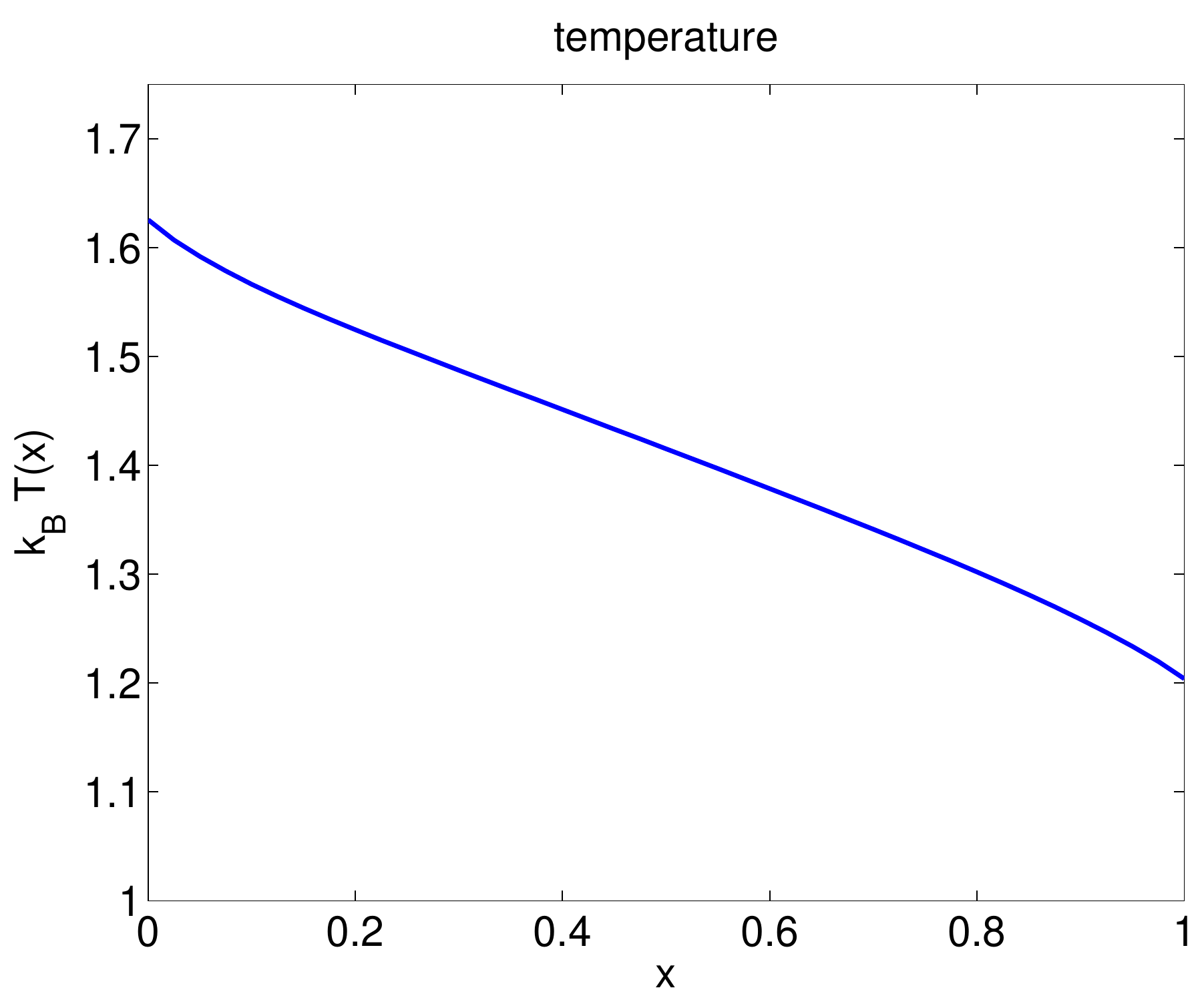}} \hspace{0.04\textwidth}
\subfloat[local entropy]{\includegraphics[width=0.3\textwidth]{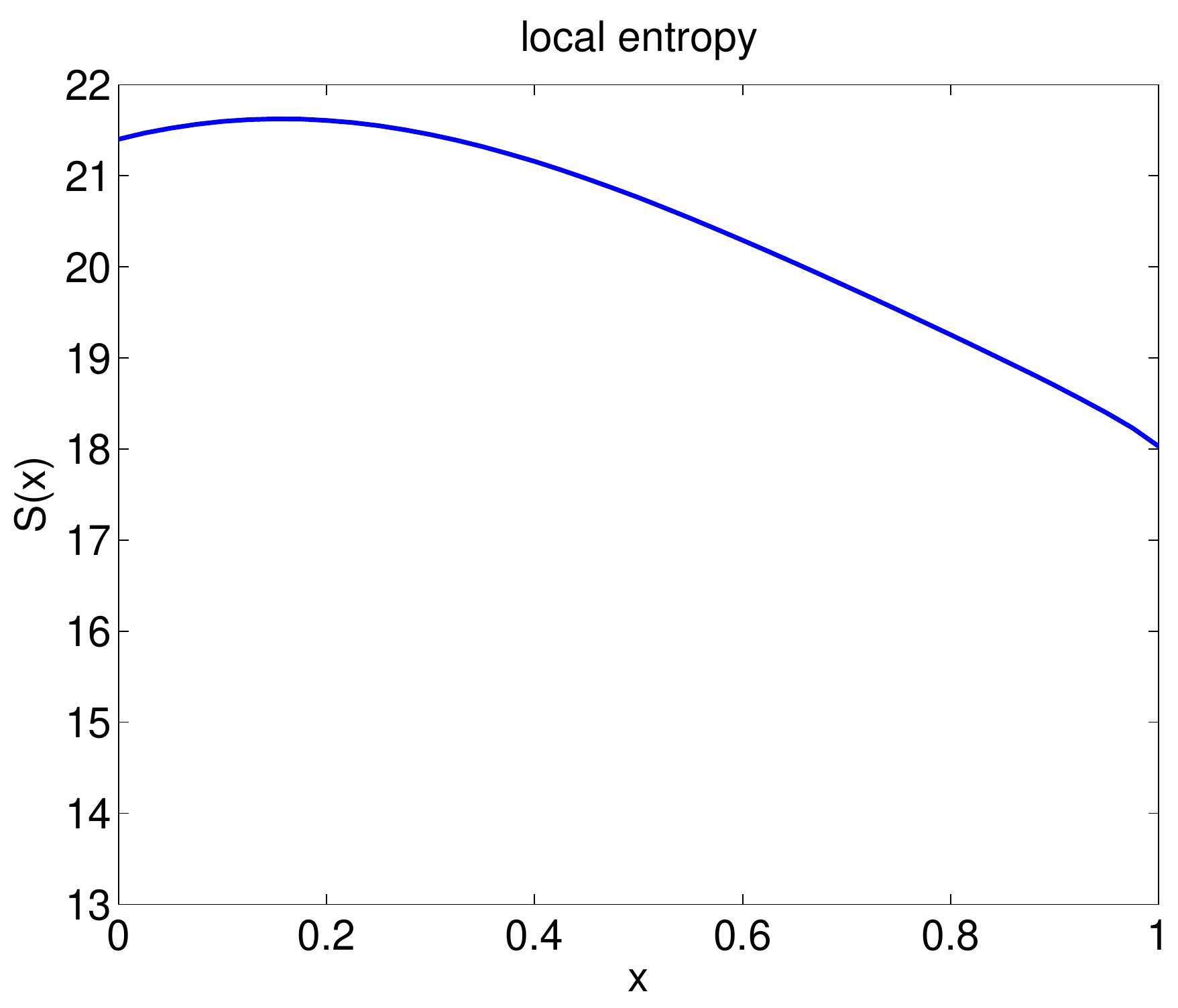}} \\
\subfloat[stationary density]{
\label{fig:Maxwell2_density}
\includegraphics[width=0.3\textwidth]{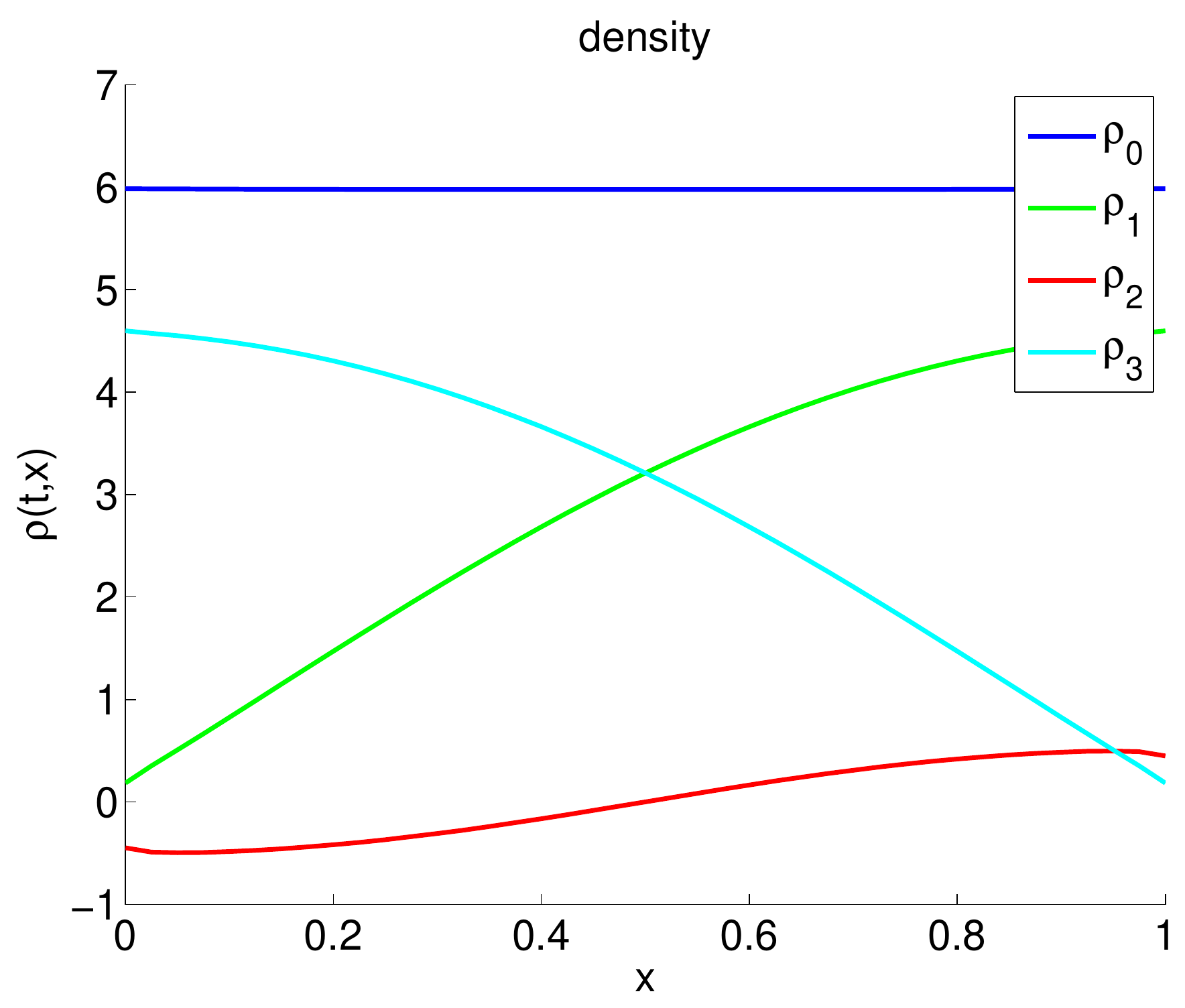}} \hspace{0.04\textwidth}
\subfloat[$k_{\mathrm{B}} T$]{
\label{fig:Maxwell2_temperature}
\includegraphics[width=0.3\textwidth]{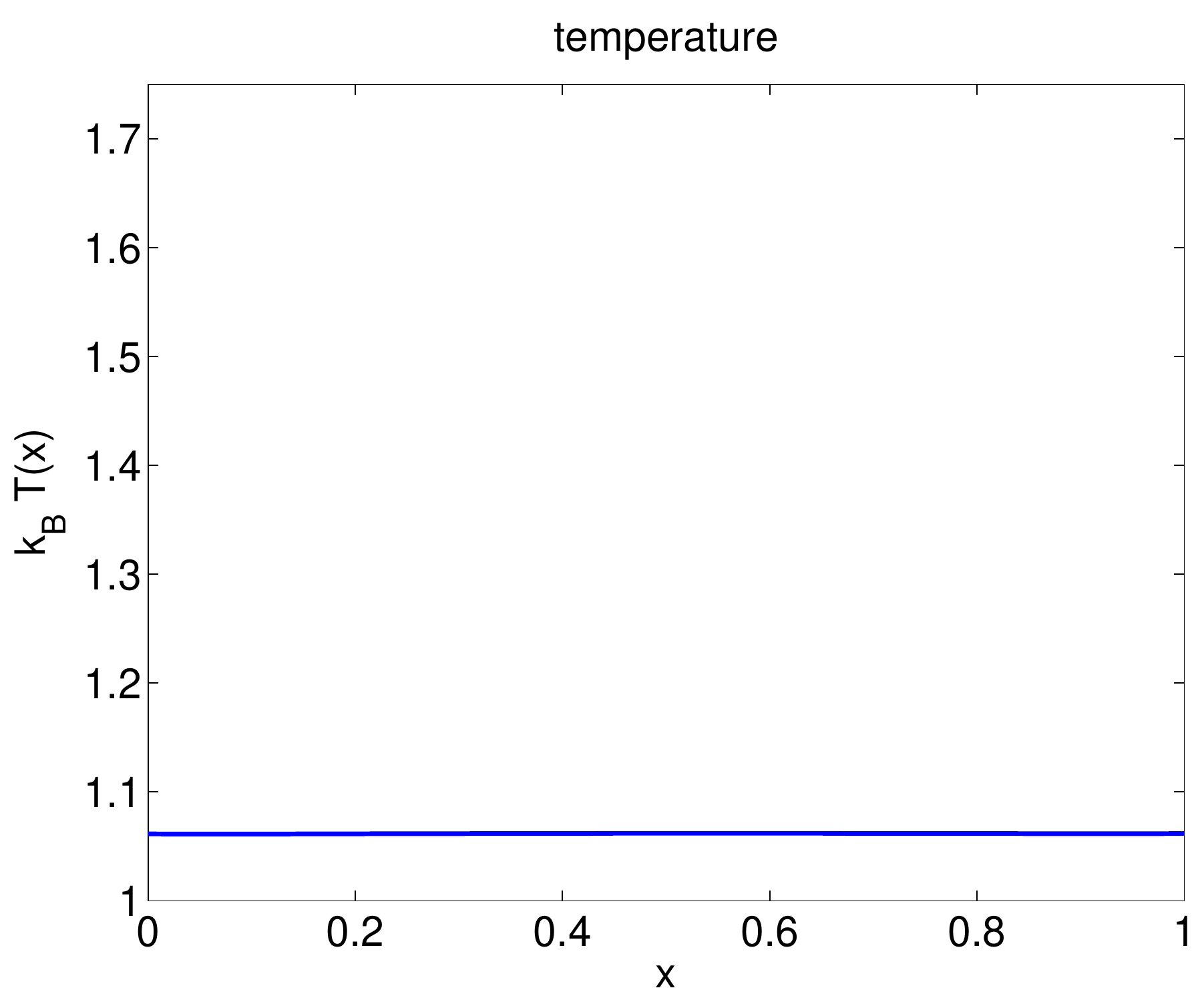}} \hspace{0.04\textwidth}
\subfloat[local entropy]{\includegraphics[width=0.3\textwidth]{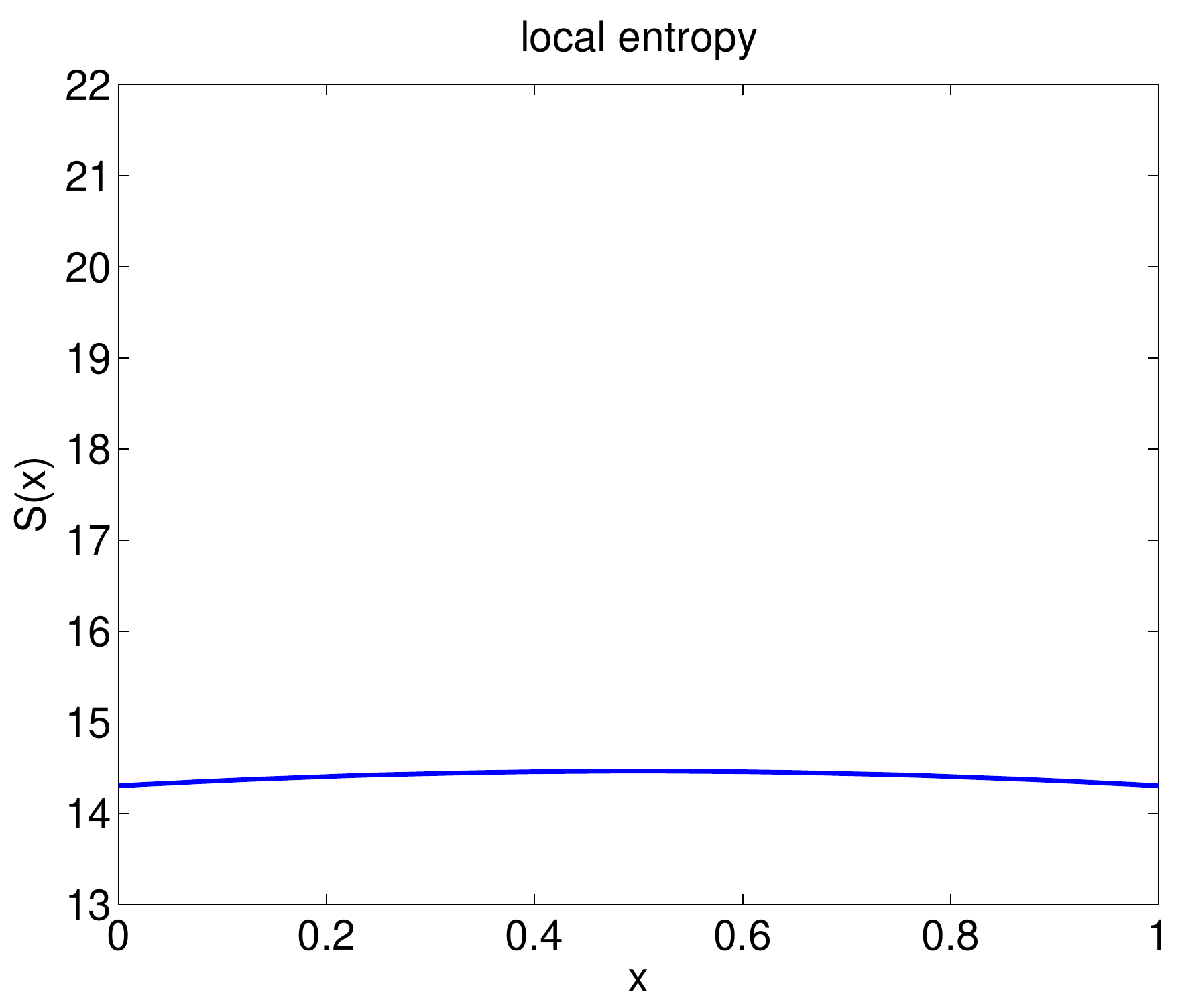}}
\caption{Components of the stationary spin density matrix, local temperature and local entropy for two simulations with Maxwell boundary conditions, with finite volume size $\Delta x = 0.025$ and $\alpha = 0.4$ (accommodation coefficient for the reflection operator). The upper row corresponds to the parameter set (i) in Table~\ref{tab:Maxwell}, and the lower row to (ii).}
\label{fig:Maxwell}
\end{figure}
The corresponding stationary states (after running the simulation up to $t = 4$) are shown in Figure~\ref{fig:Maxwell}. Somewhat surprisingly, the stationary density for case (i) with standard spin eigenbasis on the left and right contains nonzero off-diagonal entries, see Figure~\ref{fig:Maxwell1_density}. This might result from the interaction of different spin components in the collision operator, such that the local eigenbasis of the stationary state changes with spatial location. Nonzero off-diagonal entries in the stationary spin density are also observed for case (ii) in Figure~\ref{fig:Maxwell2_density}, but here this effect is certainly attributable to the rotated eigenbasis of the right Maxwell boundary condition. The stationary temperature for case (i) smoothly interpolates between the left and right boundary condition (see Figure~\ref{fig:Maxwell1_temperature}), as one might expect. The temperature for case (ii) remains constant, in accordance with the same temperature on the left and right (see Figure~\ref{fig:Maxwell2_temperature}). Finally, the local entropy tends to decrease with temperature in case (i) and remains largely unaffected by the rotation of the local eigenbasis in case (ii). Recall that the entropy \eqref{eq:entropy_def} is invariant under a change of eigenbasis. In summary, the local temperature and entropy conform with reasonable expectations, but the local spin eigenbasis of the density $\rho(t, x)$ could have hardly been predicted from the Maxwell boundary conditions.

\section{Conclusions and outlook}\label{sec:conclusion}

We have developed an efficient numerical algorithm based on spectral
Fourier discretization for the matrix-valued quantum Boltzmann
equation. The effective Hamiltonian \eqref{eq:Heff} appears only in
the matrix-valued version (since a commutator of scalars vanishes) and
consists of a principal value integral lacking microscopic energy
conservation; we have introduced a shift in the numerical grid points
\eqref{eq:theta_grid} to treat the singular part of the principal
value. The resulting algorithm exhibit spectral accuracy as
numerically confirmed in Figure~\ref{fig:conv-N}.

Our numerical simulations support the picture of fast convergence to
local equilibrium and slower global equilibration, see
Figure~\ref{fig:inhom_periodic_relentropy}. This suggests that future
work on effective hydrodynamic equations derived from the
matrix-valued quantum Boltzmann equation might be a promising
endeavor. A multiscale algorithm coupling the many-body Hubbard
model, the kinetic description of the weakly interacting Hubbard system,
and the hydrodynamic limit of the model would also be an interesting
future direction to explore.



\end{document}